\documentclass[aps,prb,twocolumn,showpacs,superscriptaddress]{revtex4}
\usepackage{amsmath}   
\usepackage{amsfonts} 
\usepackage{amssymb}
\usepackage{graphicx}   
  
\begin{document}

\title{Quantum Quenches in a $XXZ\,$ Spin Chain from a Spatially Inhomogeneous Initial State}

\author{Jarrett Lancaster}
\affiliation{New York University, Department of Physics, 4 Washington Place, New York,
NY 10003}
\author{Aditi Mitra}
\affiliation{New York University, Department of Physics, 4 Washington Place, New York,
NY 10003}
\date{\today}

\begin{abstract}
Results are presented for the nonequilibrium dynamics of a quantum $XXZ$-spin chain whose spins are 
initially arranged in a domain wall profile via the application of a magnetic field  
in the $z$-direction which is spatially varying along the chain. 
The system is driven out of equilibrium in two ways:
a). by rapidly turning off the magnetic field, b). by 
rapidly quenching the interactions at the same time as the magnetic field is turned off. The
time-evolution of the domain wall profile as well as various two-point spin correlation 
functions are studied
by the exact solution of the fermionic problem for the $XX$ chain and
via a bosonization approach and a mean-field approach for the $XXZ$ chain.
At long times the magnetization is found to equilibrate (reach the ground state value), while the 
two-point correlation functions in general do not. In particular, for quenches within
the gapless $XX$ phase, the spin correlation function transverse to the $z$ direction 
acquires a spatially inhomogeneous structure at long times whose details depend
on the initial domain wall profile.  The spatial inhomogeneity is also recovered for the
case of classical spins initially arranged in a domain wall profile and shows that the
inhomogeneities arise due to
the dephasing of transverse spin components as the domain wall broadens.  
A generalized Gibbs ensemble approach is found to be inadequate in capturing
this spatially inhomogeneous state.
\end{abstract}


\maketitle

\section{Introduction}

Recent years have seen remarkable progress in manipulating  
cold-atom systems~\cite{rev08} providing us with almost ideal realizations of
strongly correlated many-particle systems. Cold atom systems also 
have the unique feature that the interactions between
particles and the external potentials that they are subjected to are highly tunable
and can be changed rapidly in time, thus driving these systems
into highly nonequilibrium states~\cite{Weiss06}.
This property has in turn motivated a lot of theoretical activity that revolves around
studying nonequilibrium time-evolution
of strongly correlated systems arising due to a sudden
change in some parameter of the Hamiltonian referred to as a ``quantum quench''. 
The general consensus is that if the strongly correlated system
is integrable, its time-evolution from some arbitrary initial state
is highly constrained by the initial conditions
so that  these systems do not reach the ground state but instead reach 
interesting nonequilibrium time-dependent or time-independent steady states.

In this paper we study quenched dynamics in one such integrable model, namely the $XXZ$ spin chain.  
This model shows rich behavior even in its ground state~\cite{giamarchi}. For example, it 
exhibits a quantum critical point at $J_z= \pm J$ where $J$ is the exchange interactions for
the $x,y$ components of the spins, and $J_z$ is the exchange interaction for the $z$ component
of the spins. For $|J/J_z|> 1$,  the ground state is an $XX$ phase characterized by gapless 
excitations, while for $|J/J_z|< 1$, the spins are in an Ising phase where the excitation spectrum 
has a gap. Note that the special point in the gapless phase corresponding to $J_z=0$ will be
referred to as the XX-model, while the term gapless $XX$ phase will refer to the regime 
$|J/J_z|> 1$.

Nonequilibrium dynamics of the $XXZ$
model arising due to quenches has been a very active area of research
(see for example~\cite{Sengupta04,Demler09} ).
While most previous work has studied quenched dynamics in
spatially homogeneous systems, in this work we consider the nonequilibrium time-evolution arising from
quenching from an initial state which is spatially inhomogeneous. Such an initial state is created by the
application of an external magnetic field in the $z$ direction that varies in magnitude along the chain, 
changing its sign at some point so that the spins are aligned in a domain wall pattern. (See~\cite{Weld09} 
for an experimental realization of such a setup). We study nonequilibrium dynamics that arise in the
following two ways: a). by a sudden quench of the magnetic field to zero; b). 
by a sudden quench of the magnetic field to zero along with a quench in the magnitude of the  
exchange interactions. Note that the above are situations where the time evolution 
is due to a final Hamiltonian that has 
a spatially homogeneous ground state. 
We would like to understand how the system evolves in time after the quench, and whether the initial
inhomogeneity affects the properties of the system at long times. 

Some literature already exists which studies time evolution of a spin chain initially arranged in
a domain wall profile. Antal et al~\cite{antal} study the
time evolution of a sharp domain wall in the
exactly solvable $XX$ model and find ballistic broadening of the domain wall width.
Similar studies for the time-evolution of a domain wall has been carried out for
the XX model~\cite{Karevski01,Ogata02,Platini07} and the critical transverse 
spin Ising model~\cite{Karevski01,Platini05}. 
The numerical method time-dependent density matrix renormalization group (t-DMRG)
has been used to study the time-evolution of a domain wall 
both in the gapless phase as well as the gapped Ising phase of an $XXZ$ chain~\cite{tdmrg05,tdmrg09}. 
Their study reveals qualitatively different behavior in the two phases, with 
ballistic broadening of the domain wall
in the gapless phase, and more complicated non-ballistic behavior in the Ising phase. 
In addition conformal 
field theory methods~\cite{Calabrese08} have been used to study both the time-evolution 
of the domain wall 
profile as well as two-point correlation functions for the
Ising chain. Studies also exist
on the time-evolution of other quantities such as the 
entanglement entropy after a quench from a spatially 
inhomogeneous initial state~\cite{Eisler09,localquench}. 
Thus what is lacking in the literature is the study of 
the time evolution of two-point correlation functions for the $XXZ$ chain
from an initial inhomogeneous state. 
The advantage 
of studying these quantities is that they are often more sensitive to nonequilibrium initial
conditions than averages of local quantities such as the magnetization. 

In this paper we extend previous results by studying 
the time evolution of {\sl both} the local magnetization as well
as two point correlation functions related to the longitudinal (along
$z$ direction) and the transverse spin correlation function
for the $XXZ$ chain. 
In addition we use a variety
of theoretical methods such as bosonization, exact solution of the fermionic problem as well as
a classical treatment of spins which allows us to compare the relative merits of these approximations.
Consistent with
previous results, we find that the magnetization always equilibrates (i.e,
reaches the ground state value). In addition the details of how the
domain wall profile evolves in time depend on whether the system after the quench is in the gapless 
phase or the gapped Ising phase~\cite{tdmrg05,tdmrg09}. 
For example in the gapless phase the domain wall is found to spread out
ballistically, however in the gapped phase the dynamics is much more complicated showing oscillations and
revivals at short time scales. 

In studying the time-evolution of two-point correlation functions
we obtain the interesting new result that the transverse spin-correlation function shows a lack
of equilibration reaching a nonequilibrium steady state that is also spatially
inhomogeneous. The length scale of the spatial inhomogeneity is found to 
depend on the details of the initial domain wall profile and is found to arise due to
the dephasing of transverse spin components as the domain wall broadens. This result is
qualitatively different from those obtained for the Ising chain initially arranged in
a domain wall profile where no residual inhomogeneity was observed~\cite{Calabrese08}.

The paper is organized as follows. In section~\ref{XXfermions} we study the XX model which is
initially subjected to a spatially varying magnetic field so that the spins are arranged in 
a domain wall profile. We study the nonequilibrium dynamics that arises when this magnetic-field
is suddenly switched off. These results are obtained for two cases. One for an initial
external magnetic field that varies linearly in position and is equivalent to an effective 
electric field on the fermions. The second case is that of an initial magnetic field which 
has a sharp step-function profile.
In section~\ref{magXXZbos} the effect of quenching only a 
spatially varying magnetic field in the XX model is studied via a bosonization approach. The results of
section~\ref{XXfermions} and~\ref{magXXZbos} 
are found to be in qualitative agreement and in particular recover ballistic domain wall motion and
spatial inhomogeneities in
the transverse spin correlation function. A physical explanation for this inhomogeneity is provided in 
section~\ref{cleom} where the problem of 
quenching a spatially varying magnetic field is studied classically and a spin-wave pattern is 
obtained as a long time solution of a Landau-Lifshitz equation. 

In section~\ref{magintXXZ}, the effect of simultaneously quenching the
magnetic field and interactions is studied using a bosonization approach. 
In section~\ref{magintXX} results are presented 
for the case when the
quench is entirely within the gapless XX phase. The spatial inhomogeneities in the
correlation functions are recovered along with the nonequilibrium exponents obtained 
earlier in a purely homogeneous interaction quench~\cite{cazalilla1,cazalilla}.
In section~\ref{magintSC}
the effect of quenching into the gapped Ising
phase is studied via a semiclassical approach. Here we find the domain wall
dynamics to be qualitatively different, however at long times we find that all  
inhomogeneities eventually decay away. 
The latter result may very well be an artifact of the semiclassical 
approach which neglects creation of solitons. 
The results of the semiclassical approach is complemented
by a mean-field treatment in section~\ref{meanfield} where we 
address the question of whether Ising order can develop for a system initially in a domain wall state. 
Finally we conclude in section~\ref{conclusions}.

\section{Quench of a spatially varying magnetic field in the XX model: Exact solution of the
fermionic problem} \label{XXfermions}

In this section we will study the XX model with a magnetic field in the 
$z$ direction that varies linearly in position and changes sign at 
the center of the spin chain, so that the ground
state is a domain wall configuration. In subsection~\ref{DWstate} we will study the properties of this
domain wall state, and in subsequent subsections study the nonequilibrium time-evolution arising when the 
the spatially varying magnetic field is suddenly switched off. 

\subsection{Creation of domain-wall state} \label{DWstate}

Our starting point is the $XX$ Hamiltonian in a magnetic field,
\begin{equation}
\label{eq:xyham}
H_{xx} = -J\sum_{j}\left[S_{j}^{x}S_{j+1}^{x} +  
S_{j}^{y}S_{j+1}^{y} - \frac{h_j}{J}S_{j}^{z} \right],
\end{equation}
where $S_{j}^{\nu} = \frac{1}{2}\sigma_{j}^{\nu}$, $\sigma^{\nu}$ are Pauli matrices,
$h_j$ is an external magnetic field aligned along the $\hat{z}$ direction, and 
$J$ is the exchange energy. For a uniform magnetic field one commonly employs the Jordan-Wigner 
transformation~\cite{lsm,barouch1,barouch2,barouch3} to map the spin system 
to a system of spin-less non-interacting fermions which can be easily
diagonalized (for details
see Appendix~\ref{diagonol}).

To create an inhomogeneous state we will consider a linearly varying field $h_j = {\cal F} j a$,
where $a$ is a lattice spacing. Note that the ground state properties of spin-chains in 
various spatially varying magnetic fields has been studied in~\cite{DWstatics}. 
After the Jordan-Wigner transformation Eq.~\ref{eq:xyham} for $h_j = {\cal F} j a$ maps to the 
Wannier-Stark problem~\cite{Eisler09,smith,case} of electrons in a lattice subjected to a constant 
force (or electric field) ${\cal F}$,
\begin{equation}
H_{xx} = -\frac{J}{2}\sum_{j}\left[c_{j}^{\dagger}c_{j+1}+c_{j+1}^{\dagger}c_{j}\right] + \sum_{j}j
\mathcal{F}ac_{j}^{\dagger}c_{j} \label{Hxxjw}
\end{equation}
This Hamiltonian can be diagonalized as follows,
\begin{equation}
H_{xx} = \sum_{m}\epsilon_{m}\eta_{m}^{\dagger}\eta_{m},
\end{equation}
where
\begin{equation}
\epsilon_{m} = m\mathcal{F}a,\;\;\;\;\;\left(\forall \;\; m \;= \;\mbox{integer} \right),
\end{equation}
and 
\begin{eqnarray}
\eta_{m}^{\dagger} & = & \sum_{j}J_{j-m}\left(\frac{J}{\mathcal{F}a}\right)c^{\dagger}_{j}\label{eq:etaj}\\
 & = & \frac{1}{\sqrt{N}}\sum_{k}\exp\left[-ikma - i\frac{J}{\mathcal{F}a}\sin(ka)\right]c^{\dagger}_{k}
\label{eq:etaj1}
\end{eqnarray}
Above  $c_j =\frac{1}{\sqrt{N}}\sum_k c_k e^{i k j}$, $N$ 
is the total number of sites and $J_{n}(x)\,$ is a Bessel function of the first kind. 

\begin{figure}
\begin{center}
\includegraphics[totalheight=7cm,width=9cm]{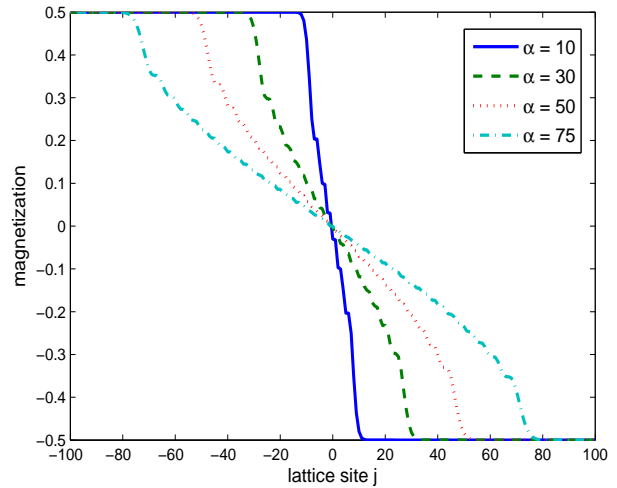}
\caption{\label{fig:initmag}(Color online) Plot of the initial magnetization for various values of 
the domain wall width $\alpha = \frac{J}{\mathcal{F}a}$.} \label{domainwall}
\end{center}
\end{figure}

We construct the ground state $\left|\Psi\right\rangle$ by including all negative energy states,   
\begin{equation}
\label{eq:initstate}
\left| \Psi \right\rangle = \prod_{m<0}\eta_{m}^{\dagger} \left| 0\right\rangle.
\end{equation}
This is a domain-wall state with a characteristic width $\frac{x}{a} \equiv 
\alpha = \frac{J}{\mathcal{F}a}$ to the left (right) of which all spins are up (down). 
This can be seen by evaluating the magnetization at the position $ja$
\begin{equation}
m^{z}(j) = \left\langle \Psi|c_{j}^{\dagger}c_{j}|\Psi \right\rangle - \frac{1}{2}.
\end{equation}
which on rewriting $c_{j},c_{j}^{\dagger}$ in terms of the $\eta_{m}$ operators 
and evaluating the expectation value is found
to be
\begin{equation}
m^{z}(j) = -\frac{1}{2} + \sum_{m\geq 0}J^{2}_{j+m}(\alpha) \label{mz}
\end{equation}
Upon noting that the support of $J_{n}(x)\,$ is restricted to~\cite{hartmann} $|n|\leq x$, and using 
the identity~\cite{abramowitz} $ \sum_{n}J_{n}^{2}(x) = 1$, it is easy to see that this
indeed describes a domain-wall state of width $\alpha$. 
Eq.~\ref{mz} is also plotted in Fig.~\ref{domainwall}.
It is worth noting that the case of $\alpha = 0$ corresponds to an 
infinite ``electric field'' $\mathcal{F}$ 
that forces a sharp wall of zero width. This case was investigated by 
Antal~{\it et al}\cite{antal}. We will
extend some of their results to more general domain walls, and also study spin-spin correlation
functions. 

The basic spin correlation functions that we will compute are,
\begin{equation}
C^{\nu\nu}(j,j+n) = \left\langle S_{j}^{\nu}S_{j+n}^{\nu} \right\rangle, \;\;\;\; (\nu = x,z)
\end{equation}
where we will refer to $C^{zz}$ ($C^{xx}$) as the longitudinal 
(transverse) spin correlation function.
These may be expressed in terms of Majorana operators~\cite{lsm} 
$A_{j} = c_{j}^{\dagger} + c_{j}$, and $B_{j} = c_{j}^{\dagger} - c_{j}$ as follows,
\begin{equation}
\label{eq:czzm}
C^{zz}(j,j+n) = \frac{1}{4}\left\langle B_{j}A_{j}B_{j+n}A_{j+n} \right\rangle,
\end{equation}
and
\begin{equation}
\label{eq:cxxm}
C^{xx}(j,j+n) = \frac{1}{4}\left\langle B_{j}A_{j+1}B_{j+1} \cdots A_{j+n-1}B_{j+n-1}A_{j+n} \right\rangle,
\end{equation}
The above correlations can be evaluated by rewriting $c_{i}$ in terms of $\eta_m$ and
applying Wick's theorem. 
The basic contractions are~\cite{zeromode} 
\begin{eqnarray}
&&\left\langle B_{j}A_{j+n} \right\rangle = -\left\langle A_{j}B_{j+n}\right\rangle = 
\sum_{m>0}\left[J_{j+m}(\alpha)
J_{j+n+m}(\alpha)\right. \nonumber \\
&&\left. - J_{j-m}(\alpha)J_{j+n-m}(\alpha)\right], \\
&&\left\langle A_{j}A_{j+n} \right\rangle = - \left\langle B_{j}B_{j+n} \right\rangle = \delta_{n=0}.
\end{eqnarray}
Using some Bessel function identities, 
the mixed contraction can be simplified to
\begin{eqnarray}
&&\left\langle B_{j}A_{j+n}\right\rangle = \frac{\alpha}{2n}\left[J_{j+n}(\alpha)J_{j+1}(\alpha) - 
J_{j+n+1}(\alpha)J_{j}(\alpha) \right. \nonumber \\
&&\left. - J_{j+n}(\alpha)J_{j-1}(\alpha) + J_{j+n-1}(\alpha)J_{j}(\alpha)\right].
\end{eqnarray}
Since we encounter no contractions of the form 
$\left\langle A_{i}A_{i}\right\rangle$, the function $C^{xx}(j,j+n)\,$ may be written\cite{lsm} as a 
Toeplitz determinant which is a computationally cheap task. In the next sub-section when we 
evaluate the correlation functions after a quench
of the magnetic field, we will 
find $\left\langle A_{j}A_{j+n} \right\rangle \neq 0$ in the nonequilibrium state. 
In this case, the evaluation of $C^{xx}$ will require computing the   
square root of a determinant~\cite{barouch2}. 

\begin{figure}
\begin{center}
\includegraphics[totalheight=7cm,width=9cm]{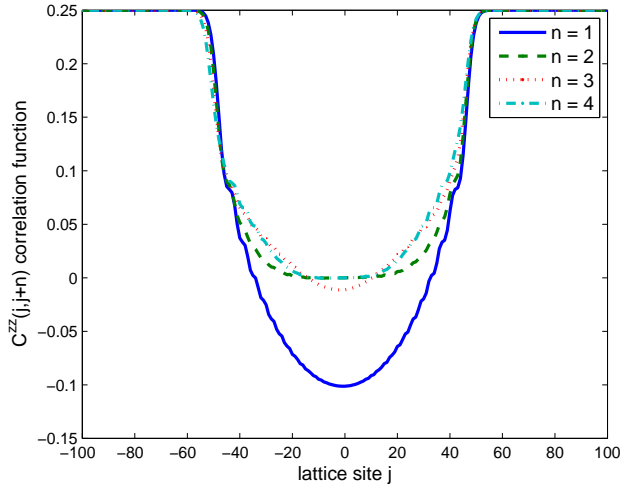}
\caption{\label{fig:initczz} (Color online) Plot of $C^{zz}(j,j+n)\,$ for $n=1,2,3,4$ and 
$\alpha =\frac{J}{{\cal F}a}= 50$.}
\end{center}
\end{figure}

\begin{figure}
\begin{center}
\includegraphics[totalheight=7cm,width=9cm]{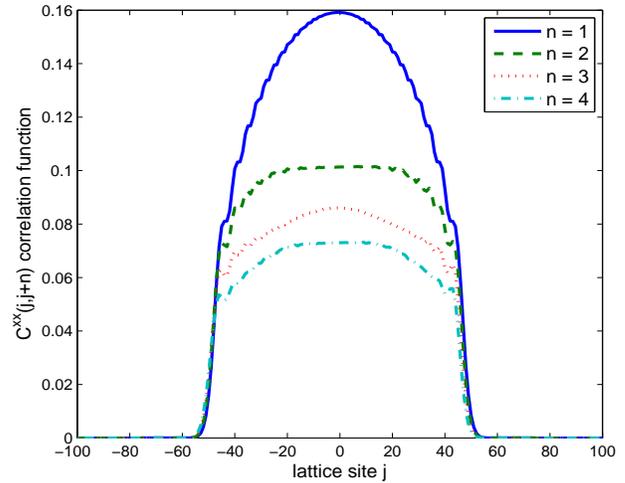}
\caption{\label{fig:initcxx} (Color online) Plot of $C^{xx}(j,j+n)\,$ for $n=1,2,3,4$ and 
$\alpha = \frac{J}{{\cal F}a} = 50$.}
\end{center}
\end{figure}

The correlation functions $C^{zz}$ and $C^{xx}$ for 
$\alpha = 50$ are given in Fig.~\ref{fig:initczz} and Fig.~\ref{fig:initcxx} 
respectively. The main point to notice here is that close to the center of the domain wall
where the external magnetic field is small, the correlations tend to mimic those in the
ground state of the homogeneous ($h_j=0$) $XX$ model. For example, 
the nearest neighbor longitudinal spin correlation function in 
the absence of a magnetic field is~\cite{barouch3} 
$C^{zz}_{eq}(n=1)=-\frac{1}{\pi^2}$. This is precisely the value that 
the $C^{zz}(j,j+1)$ correlation function takes in Fig.~\ref{fig:initczz} at the center of
the domain wall.

\subsection{Domain wall dynamics after the magnetic field quench}

We will now explore the dynamics that arises when the spatially varying magnetic field 
is suddenly switched off at $t=0$ {\sl i.e.},
$h_j(t) = \theta(-t)\mathcal{F}ja$. 
Thus at $t<0$ the many-body wavefunction of the system is Eq.~\ref{eq:initstate}, while for $t>0$, 
as shown in Appendix~\ref{diagonol}, the wave-function evolves according to the
Hamiltonian,
\begin{equation}
H_{f} = \sum_{k}\epsilon_k c_{k}^{\dagger}c_{k}
\end{equation} 
where $\epsilon_k = -J\cos(ka)$.

The time-dependent magnetization is given by
\begin{equation}
m^{z}(j,t) = \left\langle \Psi\right| c_{j}^{\dagger}(t)c_{j}(t)\left| \Psi\right\rangle - \frac{1}{2}.
\end{equation}
The time-evolution of $c_j,c_j^{\dagger}$ in terms of the $\eta_{m}$ operators is given
in Eqns.~\ref{cjtim1},~\ref{cjtim2} from which we obtain
\begin{equation}
m^{z}(j,t) = -\frac{1}{2} + \sum_{m>0}|L(j+m,t,\alpha)|^{2},
\end{equation}
where~\cite{Eisler09}
\begin{equation}
L(j+m,t,\alpha) = \int_{-\pi}^{\pi}\frac{dk}{2\pi}e^{ik(j+m)- i\epsilon_kt - i\alpha\sin k}.
\end{equation}
Evaluating the integral, we find
\begin{equation}
m^{z}(j,t) = -\frac{1}{2} + \sum_{m>0}J^{2}_{j+m}\left(\sqrt{(Jt)^{2} + \alpha^{2}}\right).
\end{equation}
This corresponds to a wall whose width $W=\sqrt{(Jt)^{2} + \alpha^{2}}$ increases
linearly and hence ballistically in time with a velocity of $Ja$ for times $t > \alpha/J$. 
Setting $\alpha = 0$, we recover 
the result of Antal {\it et al}~\cite{antal}. Note that the time-evolution of the
entanglement for general $\alpha$ has been studied in~\cite{Eisler09}.

\subsection{Correlation functions after the magnetic field quench} 

We now turn to the evaluation of the longitudinal and 
transverse correlation functions given in 
Eqns.~(\ref{eq:czzm}) and ~(\ref{eq:cxxm}) at a time $t$ after the quench. 
The basic contractions that we need are $\left\langle B_{j}(t)A_{j+n}(t)\right\rangle $ and 
$\left\langle A_{j}(t)A_{j+n}(t) \right\rangle$ for which we find the following expressions
\begin{eqnarray}
&&\left\langle B_{j}(t)A_{j+n}(t)\right\rangle  =  \nonumber \\
&&\frac{r}{2n}\left[e^{in\delta}\left(J_{j+n}(r)J_{j+1}(r) 
- J_{j+n+1}(r)J_{j}(r)\right) \right. \nonumber \\
&&\left. - e^{-in\delta}\left(J_{j+n}(r)J_{j-1}(r)-J_{j+n-1}(r)J_{j}(r)\right)\right],
\end{eqnarray}
\begin{eqnarray}
&&\left\langle A_{j}(t)A_{j+n}(t) \right\rangle  =  \nonumber \\
&&\frac{r}{2n}\left[e^{in\delta}\left(J_{j+n}(r)J_{j+1}(r) - J_{j+n+1}(r)J_{j}(r)\right) 
\right. \nonumber \\
&&\left. + e^{-in\delta}\left(J_{j+n}(r)J_{j-1}(r)-J_{j+n-1}(r)J_{j}(r)\right)\right],
\end{eqnarray}
with $r = \sqrt{(Jt)^{2} + \alpha^{2}}\,$ and $\tan\delta \equiv Jt/\alpha$. As before, we have 
$\left\langle B_{j}(t)B_{j+n}(t) \right\rangle = - \left\langle A_{j}(t)A_{j+n}(t) \right\rangle$
and $\left\langle B_{j}(t)A_{j+n}(t) \right\rangle = - \left\langle A_{j}(t)B_{j+n}(t) \right\rangle$. 

The transverse correlation function $C^{xx}(j,j+n,t)$ is given by 
\begin{equation}
|C^{xx}(j,j+n,t)| = \frac{1}{4}\sqrt{\mbox{det}{\bf C}}\label{detC},
\end{equation}
where ${\bf C}\,$ is the anti-symmetric matrix
\begin{equation}
{\bf C} = \left(\begin{array}{cc} {\bf S} & {\bf G} \\
{\bf -G} & {\bf Q}\end{array}\right),
\end{equation}
with the $n\times n\,$ sub-matrices defined by 
\begin{eqnarray}
S_{ik} = -S_{ki} 
& = & \left\langle B_{i+j-1}B_{k+j-1}\right\rangle \;\; (k>i)
\\Q_{ik} = -Q_{ki} & = & \left\langle A_{i+j}A_{k+j}\right\rangle \;\; (k>i)\\
G_{ik} & = & \left\langle B_{i+j-1}A_{k+j}\right\rangle.
\end{eqnarray}

Note that both ${\bf S}\,$ and ${\bf Q}\,$ are anti-symmetric, and thus defined 
by the elements with $k>i$, while no such restriction is placed on ${\bf G}$.

\begin{figure}
\begin{center}
\includegraphics[totalheight=6cm,width=9cm]{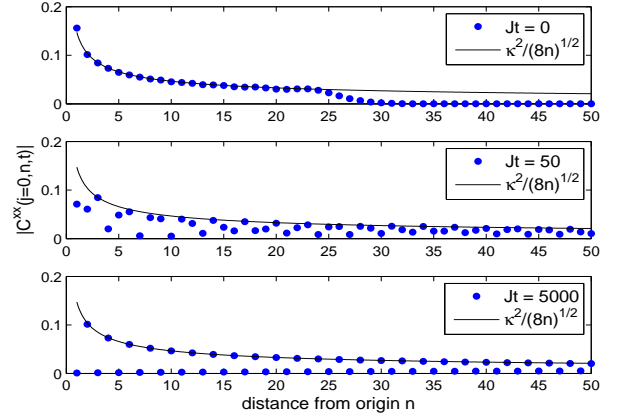}
\caption{\label{fig:finalxcorr} (Color online) Time evolution 
of $C^{xx}(j=0,n)$ correlation function for $\alpha = 25$.
Solid line is the equilibrium correlation function $C^{xx}_{eq}$.}
\end{center}
\end{figure}

The numerical evaluation of the above expression for $\alpha=25$ is shown in Fig.~\ref{fig:finalxcorr}
for three different times: $t=0$, an intermediate time and for long times where 
one finds the appearance of a spatially oscillating pattern at the scale of the Fermi-wave vector
$k_F = \pi/2$. 

In order to get some insight 
into the long time behavior of the system, we consider the limit of $r\rightarrow\infty$ while imposing
the condition $j,j+n \ll r$ on the two position indices of the correlation functions. 
This limit corresponds to a very broad domain wall which may arise either due to
a very weak electric field $\alpha \gg 1$ at arbitrary times $t$ (and hence
arbitrary $\delta$) or could arise on waiting long enough 
after a quench so that an initial narrow domain wall has had enough time to broaden. The condition on the
position indices imply that we are looking at correlations in the vicinity of the center of the
domain wall where 
the magnetization has locally equilibrated.  
Using the asymptotic expansion for the Bessel functions 
with large arguments~\cite{abramowitz}, we find
\begin{eqnarray}
\left\langle B_{j}(t)A_{j+n}(t) \right\rangle 
\sim \frac{2\cos(n\delta)}{\pi n}\sin\left(\frac{n\pi}{2}\right)
,\label{asm1}\\ 
\left\langle A_{j}(t)A_{j+n}(t) \right\rangle 
\sim \frac{2i\sin(n\delta)}{\pi n}\sin\left(\frac{n\pi}{2}\right)
\label{asm2}
\end{eqnarray}

Substituting the asymptotic expansions in Eq.~\ref{asm1},
Eq.~\ref{asm2} we find,
\begin{equation}
C^{zz}(j,j+n,t) \longrightarrow C^{zz}_{eq}(n) =-\frac{\sin^{2}\left(\frac{\pi n}{2}\right)}
{\pi^{2}n^{2}} \label{Czzres}
\end{equation}
where $C^{ab}_{eq}(n)$ is the equilibrium result ({\sl i.e.} the result in the ground state of $H_f$)
for the $ab$ correlation function~\cite{barouch3}.

Employing the asymptotic expansions in the evaluation of $C^{xx}(j,j+n)$, 
we have verified through $n\sim1000\,$ that
\begin{eqnarray}
C^{xx}(j,j+n,t) &&= C^{xx}_{eq}(n)\cos(n\delta) \\
&&\xrightarrow{t\rightarrow \infty} C^{xx}_{eq}(n) 
\cos\left(\frac{\pi n}{2}\right)\label{Cxxans}
\end{eqnarray}
where $C_{eq}^{xx}(n)\,$ is the equilibrium value of the $xx$-correlation function~\cite{lsm} 
which has the asymptotic 
form~\cite{ovchinnikov},
\begin{equation}
C^{xx}_{eq}(n) \approx \frac{1}{\sqrt{8n}}\kappa^{2},
\end{equation}
with $\log\kappa = \frac{1}{4}\int_{0}^{\infty}\frac{dt}{t}\left(e^{-4t}-\frac{1}{\cosh^{2}t}\right)$.

The appearance of these oscillations at a wave-length of $\lambda = 4$ where 
\begin{eqnarray}
C^{xx}(j,j+n,t) \xrightarrow{t\rightarrow \infty} 
C^{xx}_{eq}(n)\cos(\frac{2\pi n}{\lambda}) \label{Cxxansgen}
\end{eqnarray}
is indeed intriguing. To understand better what sets this length scale, we study 
more general initial domain walls. The one considered so far has the maximum possible
positive (negative) polarization of $S_z=\pm 1/2$ on the left (right) ends of the domain wall, 
with a width of the domain wall controlled by $\alpha=\frac{J}{{\cal F}a}$. 
Antal
et al~\cite{antal} have shown how to construct domain walls of zero width ($\alpha=0$) but
arbitrary polarization $\pm m_0$ at the two ends. 
Starting with this initial state {\sl i.e,} 
\begin{eqnarray}
S^z_j(t=0) = -m_0sgn(j) 
\end{eqnarray}
we study the time-evolution under the $XX$ model and 
extend the results of Antal {\it et al} to the study of the transverse spin correlation
function. 

\begin{figure}
\begin{center}
\includegraphics[totalheight=7cm,width=9cm]{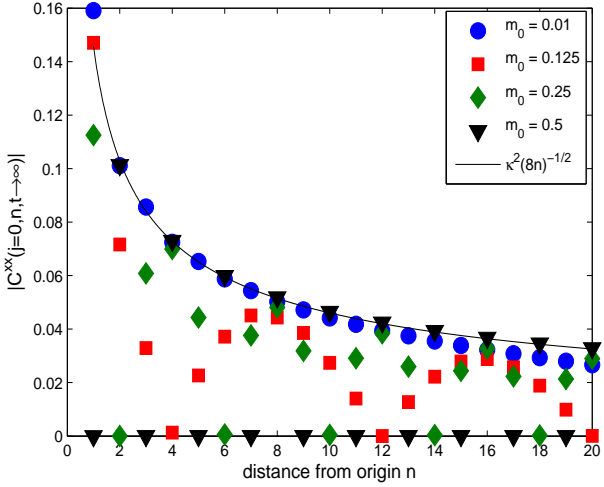}
\caption{\label{fig:cxxperiod} (Color online)
Long time behavior of $C^{xx}(j=0,n,t)\,$ for different values of the domain wall height $m_{0}$ 
defined as $\left\langle \hat{S}^{z}_{j}(t=0)\right\rangle = -sgn(j) m_{0}$. The solid line
is the equilibrium correlation function $C^{xx}_{eq}$.
}
\end{center}
\end{figure}

Details of the computation are given in Appendix~\ref{m0vary}. 
In the long time limit of $Jt\rightarrow \infty$, analytic expressions 
for contractions in Eq.~\ref{asm1},~\ref{asm2} generalize to Eq.~\ref{asmgen}. These can 
then be used to construct the determinant of $C$ (Eq.~\ref{detC}) required for the
computation of $C^{xx}$. The results for $C^{xx}$ in the 
long time limit for several different domain wall heights $m_0$
are shown in Fig.~\ref{fig:cxxperiod}. Indeed we see the appearance of a spatial oscillation
where the wavelength depends on the height $m_0$ of the domain wall as follows,
\begin{eqnarray}
\lambda = \frac{2}{m_0} \label{lamgen}
\end{eqnarray}

Thus the main results of this section on the effect of quenching a spatially varying magnetic
field in a XX model are as follows: a). An initial domain wall spreads out 
ballistically after the magnetic field quench so that the magnetization equilibrates to its
homogeneous ground state value of $m^z =0$. For a sharp domain wall, 
($\alpha =\frac{J}{{\cal F}a} \ll 1$),
the magnetization at a distance $na$ from the center of the domain wall begins to equilibrate after a time
$Jt \sim na$. This is an example of the horizon effect~\cite{cardy06,cardy08}, namely the minimal time
required for ballistically propagating excitations from the center of the domain wall
to reach the observed point;
b). Another consequence of the horizon effect is that any two point correlation
function $C^{ab}(j=0,n)$ is found to change significantly only after times~\cite{cardy08} 
$J t >  |n| $ for $\alpha \ll 1$;
c). The $C^{zz}(j,j+n)$
correlation function eventually equilibrates 
({\sl i.e.} reaches the ground state value) at long times where the approach to
equilibrium is a power-law of ${\cal O}(\frac{1}{Jt})$;
d). The $C^{xx}(j,j+n)$ 
correlation function reaches a nonequilibrium steady state at times
$min(\sqrt{\alpha^2 + (Jt)^2}, 
t {\cal F} a) \gg 1$ to a value given by Eq.~\ref{Cxxansgen} which is basically the ground state
correlation function with oscillations at a wave-length $\lambda  = \frac{2}{m_0}$ 
superimposed on it, where $m_0$ is the height of the domain wall, {\sl i.e.}, the
magnitude of the maximum polarization at its two ends. 

Note that the effect of quenching from an initial domain wall state 
was also studied in~\cite{Calabrese08}
for the critical transverse spin Ising model ($\sum_i S^x_i S^x_{i+1} + \sum_i S^z_i$)
using the methods of boundary conformal
field theory. 
Our results for the $XX$ model 
differ from those of~\cite{Calabrese08} 
in two important ways.
One is that at long times we recover a spatial spin-wave pattern that was not captured 
in~\cite{Calabrese08}. 
Secondly at long times after the quench, our two-point correlation functions continue to be
critical (i.e., have a power-law decay) whereas the results of~\cite{Calabrese08} point to
thermal behavior with exponential decay in the two-point correlation function. The differences
between these results are due to the very different universality classes and 
conservation laws of the $XX$ model and the transverse spin Ising model. For example, while
the total magnetization in the z-direction is conserved in the $XX$ model, 
it is not conserved in
the Ising model. As a result the magnetization has a nontrivial time evolution 
in the Ising model even  
for points outside the domain wall horizon, which eventually gives rise to a thermal behavior. 
In addition, as we shall show in section~\ref{cleom},
the appearance of the spin-wave pattern in the transverse spin correlation functions 
in the $XX$ model is due to 
precession of spins in an effective magnetic field created by
the initial domain wall. This precessional physics is absent for Ising spins.

We would also like to point out that the spatial oscillations found by us have 
also been observed in a different physical setting by Rigol {\it et al}~\cite{rigol04} 
who have studied inhomogeneous quantum quenches for
one-dimensional hard-core bosons. Via a Jordan-Wigner transformation hard-core
bosons in one dimension can be mapped onto
a system of spinless fermions. The physical situation considered by Rigol {\it et al}
was one where  an initial strong confining potential produces a local region
of high density. (In the spin language this would correspond to creating a pile up 
of magnetization in the center of a spin-chain). 
The time evolution
of the system when this confining potential was suddenly switched off was studied.  
The initial inhomogeneity was found to move out ballistically in two opposite directions. 
Moreover, the analog of the transverse spin correlation function 
which in the boson language corresponds to the 
off-diagonal matrix element was found to show
spatial oscillations in the two out-going lobes. (This is somewhat different from
our set-up where the
spatial oscillations appear at the center of the chain and extend over longer portions
of the chain as the domain wall broadens). 
For an initial tight confinement, Rigol {\it et al} found that the 
oscillations in the lobes appeared at precisely the 
same wave-length of $\lambda = 4$ that we find here for a domain wall of height $m_0=1/2$. 
For the case of bosons, these oscillations
had the physical interpretation of the appearance of
quasi-condensates at the wave-vector $k_F = \pm \pi/2$. A similar observation 
that an initial spatial inhomogeneity for a system of 
hard core bosons can lead to correlations in momentum space was also made 
by Gangardt {\it et al}~\cite{Pustilnik09}. 

In subsequent sections we will study the effect of quenching a spatially varying magnetic field
with and without an interaction quench in a XXZ chain using a bosonization approach. For
quenches within the gapless XX phase, results of this section, namely critical behavior 
in the two-point correlation function even
after the quench, and the presence of spatial inhomogeneities will be recovered.  
In addition we will find that the wavelength of the inhomogeneity depends on the
strength of the exchange coupling $J_z$ (and hence the Luttinger interaction
parameter).

\section{Quench of a spatially varying magnetic field in a XXZ chain: Bosonization Approach}
\label{magXXZbos}

In this section we will study a general XXZ chain which is subjected to a time-dependent
spatially varying magnetic field. The most convenient way to study this problem is via the
bosonization approach. We first review some of the notation.

\subsection{Equilibrium correlations}

The hamiltonian for the $XXZ$-spin chain in a magnetic field is
\begin{equation}
H = \sum_{j}\left[J\left(\hat{S}_{j}^{x}\hat{S}^{x}_{j+1}+\hat{S}_{j}^{y}\hat{S}^{y}_{j+1}\right) 
+ J_{z}\hat{S}_{j}^{z}\hat{S}^{z}_{j+1}- h_j S^z_j\right],
\end{equation}
The above can be mapped onto a Luttinger liquid with a back-scattering potential~\cite{giamarchi},
\begin{eqnarray}
&&H = \frac{u}{2\pi}\int dx \left[ K(\nabla \theta)^{2} + \frac{1}{K}(\nabla \phi)^{2}\right] \nonumber \\
&&+ \frac{1}{\pi}\int dx h(x)\nabla\phi- \frac{2g}{(2\pi\alpha)^{2}}\int dx \cos(4\phi(x)),
\end{eqnarray}
where $g = J_{z}a$, and the fields $\theta\,$ and $\phi\,$ are defined in terms of the boson 
creation/annihilation operators as
\begin{eqnarray}
&&\phi(x) =  -(N_{R}+N_{L})\frac{\pi x}{L} \nonumber \\
&&-\frac{i\pi}{L}\sum_{p\neq0}\left(\frac{L|p|}{2\pi}\right)^{1/2}\frac{1}{p}
e^{-\alpha|p|/2-ipx}\left(b_{p}^{\dagger} + b_{-p}\right), \\
&&\theta(x) = (N_{R}-N_{L})\frac{\pi x}{L}  \nonumber \\
&&+ \frac{i\pi}{L}\sum_{p\neq0}\left(\frac{L|p|}{2\pi}\right)^{1/2}
\frac{1}{|p|}e^{-\alpha|p|/2-ipx}\left(b_{p}^{\dagger} - b_{-p}\right). 
\end{eqnarray}
where $N_{\sigma=R,L} = \sum_{p\neq0}:\psi_{\sigma}^{\dagger}(p)
\psi_{\sigma}(p):$, and $:\cdots:\,$ denotes normal ordering of operators. In the continuum limit, 
we interpret $x = ja$ for some integer $j$ and lattice spacing $a$. The 
coupling-parameter relationships are
\begin{eqnarray}
uK & = & v_{F} = Ja\sin(k_{F}a),\\
\frac{u}{K} & = & v_{F}\left[ 1 + \frac{2J_{z}a}{\pi v_{F}}\left(1 - \cos(2k_{F}a)\right)\right].
\end{eqnarray}

For an $xy\,$ anti-ferromagnet ($J>0$), the mapping from spin language to boson language is given 
by~\cite{giamarchi}
\begin{eqnarray}
\hat{S}^{z}(x) & = & \frac{-1}{\pi}\nabla\phi(x) + \frac{(-1)^{x}}{\pi\alpha}\cos[2\phi(x)],\\
\hat{S}^{+}(x) & = & \frac{1}{\sqrt{2\pi\alpha}}\exp[-i\theta(x)]\left[(-1)^{x} + \cos[2\phi(x)]\right].
\end{eqnarray}
In equilibrium and zero magnetic fields, 
the magnetization and basic physical correlations in the gapless anti-ferromagnetic $XX$ phase are 
\begin{eqnarray}
&&\left\langle \hat{S}^{z}(x)\right\rangle = -\frac{1}{\pi}\left\langle \nabla \phi(x)\right\rangle = 0,\\
&&\left\langle \hat{S}^{z}(x+n)\hat{S}^{z}(x) \right\rangle = \nonumber \\
&&C_{1}\left(\frac{1}{n}\right)^{2} + C_{2}(-1)^{n}\left(\frac{1}{n}\right)^{2K}\\
&&\left\langle \hat{S}^{+}(x+n)\hat{S}^{-}(x)\right\rangle = \nonumber \\
&&C_{3}\left(\frac{1}{n}\right)^{2K+\frac{1}{2K}}+ C_{4}(-1)^{n}\left(\frac{1}{n}\right)^{\frac{1}{2K}}
\end{eqnarray}
where the $C_{j}\,$ are non-universal constants. 

The antiferromagnetic Hamiltonian may be transformed to a ferromagnetic Hamiltonian by the 
transformation $S^{\pm}_{j} \rightarrow (-1)^{j}S^{\pm}_{j}$, resulting in ferromagnetic operators
\begin{eqnarray}
&&\hat{S}^{z}(x)_{\mbox{\scriptsize f.m.}} = -\frac{1}{\pi}\nabla\phi(x) + \frac{(-1)^x}{\pi\alpha}
\cos[2\phi(x)],\\
&&\hat{S}^{+}(x)_{\mbox{\scriptsize f.m.}} = \frac{1}{\sqrt{2\pi\alpha}}
\exp[-i\theta(x)]\left[1 + (-1)^{x}\cos[2\phi(x)]\right]\nonumber \\
\end{eqnarray}
As a consequence, the magnetization and spin correlations in the gapless ferromagnetic $XX$ phase are,
\begin{eqnarray}
&&\left\langle \hat{S}^{z}(x)\right\rangle_{\mbox{\scriptsize f.m.}} = 
-\frac{1}{\pi}\left\langle \nabla \phi(x)
\right\rangle = 0,\\
&&\left\langle \hat{S}^{z}(x+n)\hat{S}^{z}(x) \right\rangle_{\mbox{\scriptsize f.m.}} = \nonumber \\
&&C_{1}\left(\frac{1}{n}
\right)^{2} + C_{2}(-1)^{n}\left(\frac{1}{n}\right)^{2K}\\
&&\left\langle \hat{S}^{+}(x+n)\hat{S}^{-}(x)\right\rangle_{\mbox{\scriptsize f.m.}} =  \nonumber\\ 
&&C_{3}(-1)^{n}\left(\frac{1}{n}\right)^{2K+\frac{1}{2K}} + C_{4}\left(\frac{1}{n}\right)^{\frac{1}{2K}},
\end{eqnarray}

\subsection{Magnetic field quench in the XX model ($K=1$)}\label{magXXbos}

\subsubsection{Diagonalization}
In this section we will study the time-dependent Hamiltonian
\begin{eqnarray}
H = H_i \theta(-t) + H_f \theta(t) \label{timdepH}
\end{eqnarray}
We choose a point in the $XX$ phase corresponding to the
$XX$ model $K=1$ and $u = v_{F}$. Thus,
\begin{eqnarray}
&&H_i = \frac{v_{F}}{2\pi}\int dx\left( (\nabla \theta)^{2} + (\nabla \phi)^{2} 
+ \frac{2}{v_{F}}h(x)\nabla\phi \right) \label{Hi1}\\
&&= \sum_{p\neq0}v_{F}|p|a_{p}^{\dagger}a_{p} \label{Hi2},
\end{eqnarray}
so that the initial state at $t\leq 0$ is the spatially inhomogeneous ground state of $H_i$.
The magnetic field $h(x)$ is suddenly switched off at $t=0$ so that
the system evolves according to 
\begin{eqnarray}
&&H_f = \frac{v_{F}}{2\pi}\int dx\left( (\nabla \theta)^{2} + (\nabla \phi)^{2}\right) \label{Hf1} \\
&&= \sum_{p\neq0}v_{F}|p|b_{p}^{\dagger}b_{p} \label{Hf2}
\end{eqnarray}
Since Eq.~\ref{Hi1} basically represents harmonic oscillators subjected to an electric field, it may be
easily diagonalized by the shift,
\begin{equation}
b_{p} = a_{p} + \lambda_{p}.
\end{equation}
where,
\begin{equation}
\lambda_{p} = \frac{1}{v_{F}\sqrt{2\pi|p|L}}h_{p} \label{lambda},
\end{equation}
and $h_{p}$ is the Fourier transform of $h(x)$. Taking $h(-x) = -h(x)$ (corresponding to
a magnetic field that is antisymmetric in space allowing for a domain wall solution), this gives
\begin{equation}
h_{p} = -i\int \sin(px)h(x)dx.
\end{equation}

For some time $t$ after the quench, it is straightforward to show that
\begin{eqnarray}
\phi(x,t) = e^{i H_f t} \phi(x,0)e^{-iH_f t} =\phi^{a}(x,t) + \delta\phi(x,t),\\
\theta(x,t) = e^{i H_f t} \theta(x,0)e^{-iH_f t} =\theta^{a}(x,t) + \delta\theta(x,t),
\end{eqnarray}
where 
\begin{eqnarray}
&&\phi^a(x,t) =  -\frac{i\pi}{L}\sum_{p\neq0}\left(\frac{L|p|}{2\pi}\right)^{1/2}\nonumber\\
&&\times \frac{1}{p} 
e^{-\alpha|p|/2-ipx}\left(a_{p}^{\dagger}e^{i v_F |p| t} + a_{-p}e^{-i v_F |p| t}\right),\\
&&\theta^a(x,t) = 
+ \frac{i\pi}{L}\sum_{p\neq0}\left(\frac{L|p|}{2\pi}\right)^{1/2} \nonumber\\
&&\times \frac{1}{|p|}e^{-\alpha|p|/2-ipx}\left(a_{p}^{\dagger}e^{i v_F |p| t} 
- a_{-p}e^{-i v_F |p| t}\right). 
\end{eqnarray}

Since we will 
be computing expectation values with respect to an initial state which is 
the ground state of the $a_{p}$ operators, $\phi^a,\theta^a$
will just return the equilibrium results. 
Thus the effect of the shift and hence the initial magnetic field is contained entirely in
\begin{eqnarray}
\delta\phi(x,t) & = & \frac{i}{\pi v_{F}}\int_{0}^{\infty}\frac{dp}{p}\cos[pv_{F}t]
\cos[px]h_{p}, \label{dphi}\\
\delta\theta(x,t)  & = & \frac{-i}{\pi v_{F}}\int_{0}^{\infty}\frac{dp}{p}\sin[pv_{F}t]\sin[px]h_{p}. 
\label{dtheta}
\end{eqnarray}
For any antisymmetric $h(x)$ the above reduces to,
\begin{eqnarray}
&&\delta\phi(x,t)= \frac{1}{8v_{F}}\int_{-\infty}^{\infty}dx'h(x')
\left[\mbox{sgn}[x'+z_{-}]\right. \nonumber \\
&&\left.+\mbox{sgn}[x'+z_{+}]+\mbox{sgn}[x'-z_{+}]+\mbox{sgn}[x'-z_{-}]\right],\\
&&\delta\theta(x,t)=\frac{1}{8v_{F}}\int_{-\infty}^{\infty}dx'h(x')
\left[\mbox{sgn}[x'+z_{+}]\right.\nonumber \\
&&\left.+\mbox{sgn}[x'-z_{+}]-\mbox{sgn}[x'+z_{-}]-\mbox{sgn}[x'-z_{-}]\right],
\end{eqnarray}
where $\mbox{sgn}[x]\equiv x/|x|$, and $z_{\pm} = x\pm v_{F}t$. 

The above implies that the magnetization before the quench is
\begin{eqnarray}
\left\langle \hat{S}^{z}\right\rangle = -\frac{1}{\pi}\langle \frac{\partial \phi(x)}{\partial x}\rangle
= -\frac{1}{\pi}\frac{\partial \delta\phi(x)}{\partial x}
= \frac{1}{\pi v_{F}}h(x)
\end{eqnarray}
Thus within the bosonization approach, the magnetization follows the magnetic-field profile. 

\subsubsection{Magnetization and Correlations after the quench}

The magnetization at some time $t$ after the magnetic-field quench is 
\begin{eqnarray}
&&\left\langle \hat{S}^{z}\right\rangle = -\frac{1}{\pi}\left\langle 
\partial_{x}\phi^{a}(x,t) \right\rangle 
- \frac{1}{\pi}\partial_{x}\delta\phi(x,t) \nonumber \\
&&= \frac{1}{2\pi v_{F}}\left(h(x+v_{F}t) + h(x-v_{F}t)\right) \label{eq:bosonmag}
\end{eqnarray} 
As anticipated, the domain wall spreads out with speed $v_{F}$ in both directions. 
This result is consistent
with the ballistic time-evolution found in section~\ref{XXfermions} 
from the exact solution of the fermionic problem.

Next we turn to the evaluation of the $C^{xx}$ correlation function which from the 
exact solution of the fermionic model was found to be 
$\left\langle \hat{S}^{x}_{j+n}\hat{S}^{x}_{j}\right\rangle 
\rightarrow \cos\left(\frac{2\pi n}{\lambda}\right)
C^{xx}_{eq}(n)$
where $C^{xx}_{eq}(n)\sim\frac{1}{\sqrt{n}}$ was the equilibrium correlation function, and the
wavelength of the oscillation was found to be
$\lambda = \frac{2}{m_0}$, $m_0$ being the height of the
domain wall (or the magnitude of the maximal magnetization at the ends of the domain wall).
 
In the bosonization approach,
\begin{eqnarray}
&&\left\langle \hat{S}^{+}(x+n,t)\hat{S}^{-}(x,t) \right\rangle \sim \nonumber\\
&&\exp[-i(\delta\theta(x+n,t)-\delta\theta(x,t))]
\frac{1}{\sqrt{n}}.
\end{eqnarray}
Let us suppose
\begin{eqnarray}
h(x) = h_{0}\tanh(x/\xi) \label{hdef}
\end{eqnarray}
implying 
\begin{eqnarray}
h_p = -i h_0\left(\frac{\pi \xi}{\sinh\left(\frac{\pi p\xi}{2}\right)}\right)
\label{hpdef}
\end{eqnarray}
Then we find,
\begin{eqnarray}
&&\delta\theta(x+n,t)-\delta\theta(x,t)= \nonumber \\
&&-\frac{\xi h_{0}}{2v_{F}}\log\left[\frac{\cosh[(x-v_Ft)/\xi]\cosh[(x+n+v_Ft)/\xi]}
{\cosh[(x+n-v_Ft)/\xi]\cosh[(x+v_Ft)/\xi]}\right]\nonumber \\
&&\xrightarrow{v_Ft, |v_Ft \pm x|, |v_F t \pm (x+n)|\gg \xi} 
-\frac{\xi h_{0}}{2v_{F}}\left[\frac{|x-v_Ft|}{\xi} \right. \nonumber \\
&&\left. +
\frac{|x+n+v_Ft|}{\xi}
-\frac{|x+n-v_Ft|}{\xi}-\frac{|x+v_Ft|}{\xi}\right]
\end{eqnarray}
For $v_Ft >> |x|,|x+n|$, {\sl i.e}, long after the moving domain wall front has crossed the two
observation points at $x$ and $x+n$, we find
\begin{equation}
\delta\theta(x+n,t)-\delta\theta(x,t)
\rightarrow \frac{h_{0}n}{v_{F}}.
\end{equation}
so that 
\begin{equation}
\left\langle \hat{S}^{+}(x+{n})
\hat{S}^{-}(x)\right\rangle \sim e^{i{n}h_{0}/v_{F}}\left(\frac{1}{\sqrt{n}}\right)
\label{S+-bos}.
\end{equation}

Thus we find that for an initial domain wall that has the profile
\begin{equation}
\left\langle S^{z}(x=\pm \infty) \right\rangle \longrightarrow \pm \frac{h_{0}}{\pi v_{F}}
= \pm m_0
\end{equation}
the transverse spin correlation function at long times do not equilibrate but acquire
a spatial oscillation at the wavelength 
\begin{eqnarray}
\lambda = \frac{2\pi v_F}{h_0} = \frac{2}{m_0} 
\end{eqnarray}
This result is identical to that obtained from the
exact solution of the fermionic problem (Eqns.~\ref{Cxxansgen},~\ref{lamgen}). 

Note that the wavelength of the spatial oscillation is set by
the height of the domain wall $m_0$ and is independent of the width
$\xi$ (or $\alpha = \frac{J}{{\cal F}a}$ in the fermionic problem). 
For the case of a linearly varying magnetic field studied in the
previous section, the domain wall magnetization 
approached its maximum possible value
$\mp\frac{1}{2}\,$ at its two ends. This in the boson language corresponds to  
\begin{equation}
\frac{h_{0}}{v_{F}} \rightarrow -\frac{\pi}{2a},
\end{equation}
so that writing $n = \bar{n} a$ in Eq.~\ref{S+-bos}, where $\bar{n}$ is an integer, 
\begin{equation}
\left\langle S^{+}(x+\bar{n}a)S^{-}(x)\right\rangle \rightarrow e
^{i\pi \bar{n}/2}\left(\frac{1}{\sqrt{\bar{n}}}\right),
\end{equation}
giving us precisely the observed behavior of Eq.~\ref{Cxxans}.

For the evaluation of the $\langle S^z S^z\rangle$ correlation function, we need to evaluate
$\exp\left[i\delta\phi(x+n,t)-i\delta\phi(x,t)\right]$. 
We find,
\begin{eqnarray}
&&\delta\phi(x+n,t) - \delta\phi(x,t) = \nonumber \\
&&\frac{-h_{0}\xi}{2v_{F}}\log\left[\frac{\cosh\left[\frac{x+n+v_Ft}{\xi}\right]
\cosh\left[\frac{x+n-v_Ft}{\xi}\right]}{\cosh\left[\frac{x+v_Ft}{\xi}\right]
\cosh\left[\frac{x-v_Ft}{\xi}\right]}\right]\\
&&\xrightarrow{v_Ft\gg \xi, |x|,|x+n|} 0
\nonumber
\end{eqnarray}
The above result together with the fact that 
\begin{eqnarray}
&&\partial_{x}\delta\phi(x+n,t)\partial_{x}\delta\phi(x,t) = \nonumber\\
&&\frac{1}{4\pi v_{F}^{2}}\left(h(x+v_{F}t) + h(x-v_{F}t)
\right)\nonumber \\
&&\times \left(h(x+n+v_{F}t) + h(x+n-v_{F}t)\right).
\end{eqnarray}
implies that for $v_Ft\gg \xi,|x|,|x+n|$, the $C^{zz}$ correlation function equilibrates which is in
agreement with the results in section~\ref{XXfermions}.

\begin{figure}
\begin{center}
\includegraphics[totalheight=7cm,width=9cm]{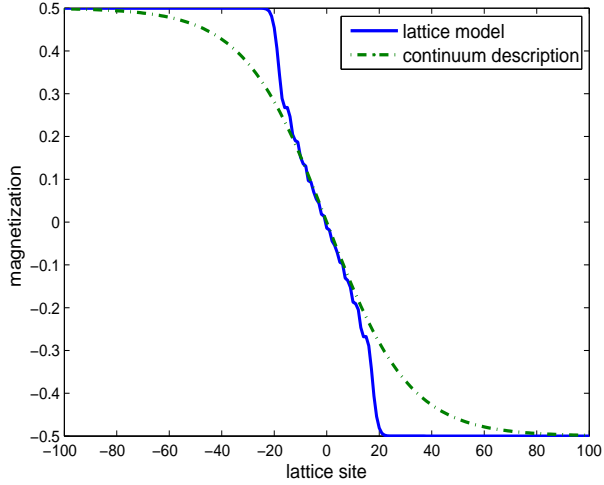}
\caption{\label{fig:comparison} (Color online) Plot of the magnetization obtained from
the bosonization approach with $h(x)=h_0\tanh{x/\xi}$, 
and from the fermionic lattice model with $\alpha = 20$.}
\end{center}
\end{figure}

To summarize the results of this section: a). Within the bosonization
approach, an applied magnetic-field 
produces a spin pattern that follows the local magnetic field. b). After a quench, the
domain wall spreads out ballistically. The magnetization equilibrates (in this case vanishes to
zero everywhere).
c). The $C^{zz}$ correlation function  equilibrates, whereas the $C^{xx}$
correlation function  acquires spatial oscillations at wavelength 
$\lambda = \frac{2}{m_0}$
where $m_0$ is the height of the initial domain wall defined as $S_z(\pm \infty) = \pm m_0$.
These results are in agreement with the exact results of section~\ref{XXfermions}
obtained from solving the fermion lattice problem. Thus bosonization is good at capturing 
the behavior at long distances
and times, even for the nonequilibrium problem. As expected it misses some of the details
of the short distance physics both in the static properties (as shown in Fig.~\ref{fig:comparison}),
as well as during the time-evolution such as the existence of 
a complex internal structure in the propagating domain wall
front~\cite{Hunyadi04}.

In the next section we provide an explanation for the inhomogeneities observed in the
transverse spin correlation function by solving the classical
equations of motion for the spins.

\section{Magnetic field quench in a classical $XX$ spin chain: 
Formation of a transverse spin-wave pattern} 
\label{cleom}

The Hamiltonian in Eq.~\ref{eq:xyham} implies the following equations of motion for the spins
\begin{eqnarray}
&&\frac{d S^z_j}{dt} = -J\left[S^y_j\left(S^x_{j+1} + S^x_{j-1}\right)- 
S^x_j\left(S^y_{j+1} + S^y_{j-1}\right)\right]\nonumber\\
\label{Szeom}\\
&&\frac{d S^x_j}{dt} = -J S^z_j\left(S^y_{j+1} + S^y_{j-1}\right) - h_j S^y_j \label{Sxeom}\\
&&\frac{d S^y_j}{dt} = J S^z_j\left(S^x_{j+1} + S^x_{j-1}\right) + h_j S^x_j \label{Syeom}
\end{eqnarray}
We will treat the spins as classical variables, and solve the above 
equations for a sudden quench 
of a spatially inhomogeneous magnetic field, $h_j(t) = h_j\theta(-t)$.
First notice that for $t\leq0$, after performing the course-graining $S^a_j = S^a_{j+1}$, 
the following spin profile satisfy the equations of motion,
\begin{eqnarray}
S^z_j = -\frac{h_j}{2J}\\
S^x_j = \sqrt{S^2 - (S^z_j)^2}\\
S^y_j =0
\end{eqnarray}

\begin{figure}
\begin{center}
\includegraphics[totalheight=7cm,width=9cm]{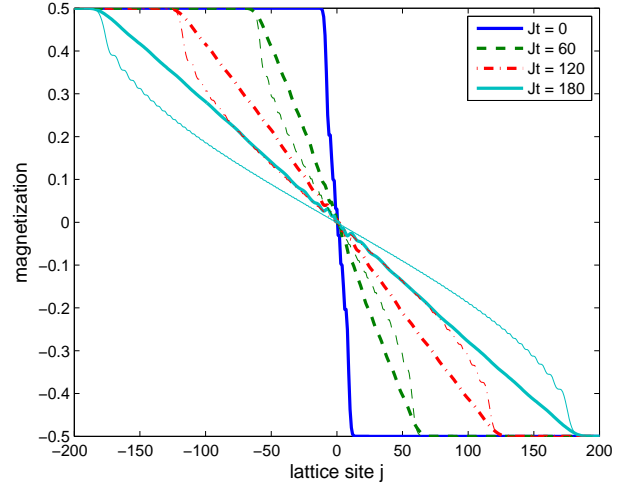}
\caption{\label{semicl} (Color online) Thick line: Time evolution of the magnetization
from the solution of the classical Landau-Lifshitz equation assuming that $S_z(t\leq0)$
is given by Eq.~\ref{mz}. 
Thin lines: Time evolution from the
exact solution of the quantum problem.}
\end{center}
\end{figure}

With the above as an initial condition, the dynamics for $t>0$ can be easily studied numerically
and we find a ballistic spreading out of the domain wall profile (see Fig.~\ref{semicl}), and the
appearance of a spin-wave pattern. 
The latter result, {\sl i.e.},
the appearance of an inhomogeneous pattern at long times can be verified 
analytically rather easily as follows.
Suppose, $h_j(t) = -h_0\tanh\frac{x}{\xi}\theta(-t)$. Then for $t>0$, 
if we assume a ballistic broadening of the
domain wall,
\begin{eqnarray}
S^z_j(t) = \frac{h_0}{2J}\left[\tanh(\frac{ja-vt}{\xi}) + \tanh(\frac{ja+vt}{\xi})\right] 
\label{Szguess}
\end{eqnarray} 
Writing Eq.~\ref{Sxeom},~\ref{Syeom} in terms of $S^{\pm} = S^x \pm i S^y$, and performing a 
course-graining, one obtains the following equation for $S^+(x = ja)$
\begin{eqnarray}
\frac{d S^+(x,t)}{dt} = 2i J S^z(x,t) S^+(x,t) \label{S+eom}
\end{eqnarray}
Using Eq.~\ref{Szguess}, Eq.~\ref{S+eom} may be easily solved to obtain
\begin{eqnarray}
S^+(x,t) = S^+(x,0)\exp\left[\frac{ih_0 \xi}{2 v}\left(\ln\frac{\cosh(\frac{vt + x}{\xi}}
{\cosh(\frac{vt-x}{\xi}}\right)\right]
\end{eqnarray}
Thus for $vt \gg |x|$ we find the spin-wave pattern
\begin{eqnarray}
S^+(x,t) \rightarrow S^+(x,0) e^{i h_0 x/v}  \label{S+class}
\end{eqnarray}
in agreement with the results of the previous sections. 

Thus the appearance of the spatial inhomogeneity may be understood as follows. Right 
after the magnetic field is switched off,
the spins in the XY plane begin to precess due to a non-zero magnetization in the
z-direction. However this precession only lasts for as long as it takes the domain wall profile to 
flatten out to zero. Since this time is different for spins located at different spatial positions,
the spins locally dephase with respect to each other and arrange themselves in a spin-wave pattern. 

\section{Quench of a spatially varying magnetic field and interactions
in a XXZ chain: Bosonization Approach} \label{magintXXZ}

\subsection{Quench within the gapless XX phase} \label{magintXX}

We will now turn our attention to the case where as the magnetic field is switched off
at $t=0$, the interactions
are also turned on so that the Luttinger parameter is $K = \theta(-t) + K \theta(t)$. We will first
consider the case where the ground state both before and after the quench corresponds to a gapless 
XX phase, so that the $\cos(\beta\phi)$ term can be neglected.

\subsubsection{Bogoliubov rotation}

We wish to begin with a non-interacting system in an inhomogeneous magnetic field,
\begin{eqnarray}
&&H_{i} = \frac{v_{F}}{2\pi}\int dx\left( (\nabla \phi)^{2} + (\nabla\theta)^{2} 
+\frac{2}{v_{F}}h(x)\nabla\phi\right) \\
&&= \sum_{p\neq0}v_{F}|p|a_{p}^{\dagger}a_{p},
\end{eqnarray}
and quench to zero field, while turning on interactions so that 
\begin{eqnarray}
&&H_{f} = \frac{u}{2\pi}\int dx\left( \frac{1}{K}(\nabla \phi)^{2} + K(\nabla\theta)^{2} \right) \\
&&= \sum_{p\neq0}u|p|\eta_{p}^{\dagger}\eta_{p}.
\end{eqnarray}
The $\eta_{p}\,$ operators have a simple time evolution and may be related to the $b_{p}\,$ 
operators by a Bogoliubov rotation~\cite{cazalilla1,cazalilla},
\begin{eqnarray}
\eta_{p} & = & \cosh\beta b_{p} + \sinh\beta b_{-p}^{\dagger},\\
\eta_{p}^{\dagger} & = & \cosh\beta b_{p}^{\dagger} + \sinh\beta b_{-p}.
\end{eqnarray}
where $e^{-2\beta} = K$.
We define,
\begin{eqnarray}
f(p,t) & = & \cos(u|p|t) - i\sin(u|p|t)\cosh(2\beta),\\
g(p,t) & = & i\sin(u|p|t)\sinh(2\beta).
\end{eqnarray}
and $\alpha_{\pm} \equiv f(p,t) \pm g(p,t)$. 
Performing the shift $b_{p} = a_{p} + \lambda_{p}$ as before, we write the $\phi$ and $\theta$ 
fields as
$(\phi, \theta) = (\phi_{\beta}^{a}, \theta_{\beta}^{a}) + \delta(\phi_{\beta}, \theta_{\beta})$, where
\begin{eqnarray}
&&\phi_{\beta}^{a}(x,t) = \frac{-i\pi}{L}\sum_{p\neq0}\left(\frac{L|p|}{2\pi}\right)^{1/2}\frac{1}{p}
e^{-\alpha|p|/2-ipx}\left(\alpha_{+}^{*}a_{p}^{\dagger} \right. \nonumber \\
&&\left. + \alpha_{+}a_{-p}\right)\\
&&\theta_{\beta}^{a}(x,t) = \frac{i\pi}{L}\sum_{p\neq0}\left(\frac{L|p|}{2\pi}\right)^{1/2}\frac{1}{|p|}
e^{-\alpha|p|/2-ipx}\left(\alpha_{-}^{*}a_{p}^{\dagger} \right. \nonumber\\
&&\left. - \alpha_{-}a_{-p} \right),
\end{eqnarray}
The $\delta(\phi,\theta)$ arise due to the spatially varying magnetic field and are given by, 
\begin{eqnarray}
\delta\phi_{\beta}(x,t) & = & \frac{u}{v_{F}}\delta\phi(x,t,\beta=0,v_{F} \rightarrow u),\\
\delta\theta_{\beta}(x,t) & = & \frac{u e^{2\beta}}{v_{F}}\delta\theta(x,t,\beta=0,v_{F}\rightarrow u),
\end{eqnarray}
where $\delta\phi(x,t,\beta=0,v_{F}),\delta\theta(x,t,\beta=0,v_{F})$
are given in Eq.~\ref{dphi} and~\ref{dtheta} respectively.

\subsubsection{Magnetization after the quench}

It is straightforward to show that the magnetization at a time $t$ after the quench is
\begin{eqnarray}
S_z(x,t) = -\frac{1}{\pi}\frac{\partial \delta\phi(x)}{\partial x} = \frac{1}{2\pi v_F}\left[h(x+ut)
+ h(x-ut)\right]\nonumber\\
\end{eqnarray}
Thus even with the interaction quench, the domain wall spreads out ballistically, but with 
the renormalized velocity $u = v_F/K$.

\subsubsection{Basic non-equilibrium correlations}

To compute the correlation functions, some basic expressions we need are 
\begin{eqnarray}
G_{\phi\phi} = \left\langle [\phi^{a}_{\beta}(x,t) - \phi^{a}_{\beta}(x',t)]^{2}\right\rangle\\
G_{\theta\theta} = \left\langle [\theta^{a}_{\beta}(x,t) - \theta^{a}_{\beta}(x',t)]^{2}\right\rangle
\end{eqnarray}
Note that $G_{\theta\theta}\,$ is related to $G_{\phi\phi}$ by $K\rightarrow\frac{1}{K}$ 
the computation of which is identical to that already presented in~\cite{cazalilla1,cazalilla}. 
For completeness we present the results below,
\begin{eqnarray}
&&G_{\phi\phi} = 
\int_{0}^{\infty}\frac{dp}{p}\left[\cos^{2}(utp) + e^{-4\beta}\sin^{2}(utp)\right]\nonumber \\
&&\times (1-\cos[p(x-x')])
e^{-\alpha p}
\end{eqnarray}
which eventually gives
\begin{eqnarray}
&&G_{\phi\phi} = G_{\phi\phi}^{(0)} + \left(\frac{K^{2}-1}{8}\right)\times \nonumber\\
&&\log\left[\frac{[\alpha^{2}+(2ut)^{2}]^{2}
[\alpha^{2}+x^{2}]^{2}}{\alpha^{4}[\alpha^{2}+(2ut + x)^{2}][\alpha^{2} + (x-2ut)^{2}]}\right] \\
&&G_{\theta\theta} = G_{\theta\theta}^{(0)} + \left(\frac{K^{-2}-1}{8}\right) \times \nonumber \\
&&\log\left[\frac{[\alpha^{2}+(2ut)^{2}]^{2}
[\alpha^{2}+x^{2}]^{2}}{\alpha^{4}[\alpha^{2}+(2ut + x)^{2}][\alpha^{2} + (x-2ut)^{2}]}\right].
\end{eqnarray}
where  
$G_{\phi\phi}^{(0)} = \frac{K}{2}\log\left[\frac{x^{2}+\alpha^{2}}{\alpha^{2}}\right]$,
$G_{\theta\theta}^{(0)}=\frac{1}{2K}\log\left[\frac{x^{2}+\alpha^{2}}{\alpha^{2}}\right]$.

\subsubsection{Spin correlations after the quench}

Taking the $2ut >> x\,$ limit, we can easily obtain the
correlator asymptotics. The memory of the initial domain wall profile 
appears as oscillatory pre-factors of the form $\exp[2i(\delta\phi(x+n)-\delta\phi(x))]$ and terms such as 
$\nabla\delta\phi(x+n)\nabla\delta\phi(x)$. For the specific case of $h(x) = h_{0}\tanh(x/\xi)$ we find
at long times,
\begin{eqnarray}
&&\left\langle \hat{S}^{z}(x+n)\hat{S}^{z}(x)\right\rangle = f(x+n,t)f(x,t) \nonumber\\
&&+ \frac{C'_{1}}{n^{2}} 
+ \frac{C'_{2}(-1)^{n}}{n^{K^{2}+1}} \\
&&\left\langle \hat{S}^{+}(x+n)\hat{S}^{-}(x)\right\rangle= 
\frac{C'_{3}e^{\frac{ih_{0}n}{v_{F}K}}}{n^{1+K^{2} + \frac{1+K^{-2}}{4}}} \nonumber\\
&&+\frac{C'_{4}(-1)^{n}e^{\frac{ih_{0}n}{v_{F}K}}}{n^{\frac{1+K^{-2}}{4}}}\label{S+-int}
\end{eqnarray}
where
\begin{eqnarray}
f(x,t) = \frac{1}{2v_{F}}\left(h(x+ut) + h(x-ut)\right)
\end{eqnarray}
The case of the pure interaction quench ($h_0=0$) was
studied in~\cite{cazalilla1,cazalilla} where the above anomalous power-laws
corresponding to an algebraic decay which is faster 
than in equilibrium (since $K^{2}+1\geq2K\,$ for all real $K$)
were obtained. 
From Eq.~\ref{S+-int} we find that the effect of an initial domain wall profile is to
superimpose spatial oscillations onto this decay, where the
wavelength of the oscillations depends on the interaction parameter $K$
and is found to be
\begin{eqnarray}
\lambda= \frac{2 \pi v_F K}{h_0} = \frac{2 K}{m_0} \label{lamint}
\end{eqnarray} 
As expected, increasing the exchange interaction $J_z$ increases the effective magnetic
field seen by the precessing spins, thus decreasing the wave-length of the spin wave
pattern.

\subsubsection{Generalized Gibbs Ensemble}

Rigol~{\it et al} have made the interesting proposal that the long-time properties of 
out of equilibrium integrable systems may be captured by a generalized Gibbs ensemble
that enforces the constraints imposed by the integrals of motion~\cite{rigol07}. 
This approach has been applied with success by Iucci and Cazalilla~\cite{cazalilla1,cazalilla} 
who studied homogeneous interaction quenches in models similar to those studied in this paper.
It has also been successfully applied to models of free bosons~\cite{Cardy07}, and for some
general family of integrable models~\cite{Mussardo09}.
An interesting question has been to identify situations in which the Gibbs ensemble argument
might break down~\cite{Barthel08,Biroli09}. 
We find that for an inhomogeneous quantum quench starting from
an initial domain wall state, the Gibbs ensemble argument
cannot capture the spatial inhomogeneities at long times after the quench. In this section we review
the Gibbs ensemble argument and show why this does not work for our case. 

Consider a quantum quench of the form $H = H_{i}\theta(-t) + H_{f}\theta(t)$,
so that for $t\leq 0$ the system is in the ground state of $H_i$, whereas at $t>0$, the wave-function
begins to evolve according to $H_f$. Let us further assume that both $H_i$ and $H_f$ can be diagonalized 
\begin{eqnarray}
H_{i} = \sum_{p}\epsilon^a_{p}a_{p}^{\dagger}a_{p}\\
H_f =  \sum_{p}\epsilon^b_{p}b_{p}^{\dagger}b_{p}
\end{eqnarray}
and that the two Hilbert spaces may be related to each other via
a canonical transformation. For the case of bosons a general canonical
transformation is of the form, 
\begin{eqnarray}
b_{p}  =  \cosh\Theta_p a_{p} + \sinh\Theta_p a_{-p}^{\dagger} + \lambda_{p}
\end{eqnarray}
where $\lambda_p$ is a linear shift and $\Theta_p$ is a rotation angle . 

The essential idea behind predicting the long-time
behavior is that an integrable model such as $H_f$ has a conserved quantity   
for every degree of freedom. For the example given above, the conserved quantity
is trivially given by 
$I_{p}=b_{p}^{\dagger}b_{p}$. Thus during the time-evolution $\langle b^{\dagger}_{p}
b_p\rangle$ should be conserved.
Since the initial state is the ground state of $H_i$ for which 
$\left\langle a_{p}^{\dagger}a_{p}
\right\rangle =0\,$ for $p\neq 0$, it follows that the Gibbs' ensemble should
be characterized by the distribution function
\begin{equation}
\left\langle b_{p}^{\dagger}b_{p}\right\rangle = |\lambda_{p}|^{2} + \sinh^{2}\Theta_{p}.
\end{equation}

For the particular models studied here, we find the initial spatial
inhomogeneity is captured by the linear shift $\lambda_p$ 
(given by Eq.~\ref{lambda}), while the angle $\Theta_p$ captures the homogeneous
changes in the interaction parameter. Since in our example $|\lambda_p|^2$ vanishes
in the thermodynamic limit, any effect of the initial spatial inhomogeneity is
lost. Thus while the Gibbs ensemble
captures the effect of the rotation, and therefore correctly predicts  
the new power law exponents of~\cite{cazalilla1}, it misses the spin-wave pattern
at long times. In particular the Gibbs ensemble prediction for the transverse spin
correlation function is given by 
\begin{eqnarray}
&&\left\langle \hat{S}^{+}(x+n)\hat{S}^{-}(x)\right\rangle= 
\frac{C'_{3}}{n^{1+K^{2} + \frac{1+K^{-2}}{4}}} \nonumber\\
&&+\frac{C'_{4}(-1)^{n}}{n^{\frac{1+K^{-2}}{4}}}\label{Sgibbs}
\end{eqnarray}
which differs from the correct answer (Eq.~\ref{S+-int}) by the absence of 
the spatially oscillating factors $e^{\frac{ih_{0}n}{v_{F}K}}$.

\subsection{Magnetic field and Interaction quench into the gapped Ising phase: Semiclassical treatment}
\label{magintSC}

In this section we will study the effect of quenching the external magnetic field 
and the interactions into the massive Ising phase where the
$\cos\phi$ term is strongly relevant. In order to study the time evolution, 
we will employ a semi-classical approximation which corresponds to replacing  
$\cos\phi \sim 1-\frac{\phi^2}{2}$. 
Thus for this case the initial wave-function is the ground state of
\begin{eqnarray}
H_{i} &= & \frac{v_{F}}{2\pi}\int dx \left[(\nabla\phi)^{2} + (\nabla\theta)^{2} + 
\frac{2}{v_{F}}h(x)\nabla\phi\right],\\
&=& \sum_p v_F |p| a_{p}^{\dagger} a_p
\end{eqnarray}
while for $t>0$ the wave-function evolves according to
\begin{eqnarray}
H_{f} & = & \frac{u}{2\pi}\int dx \left[(\nabla\tilde{\phi})^{2} + (\nabla\tilde{\theta})^{2} 
+ \frac{m^{2}}{u^{2}}\tilde{\phi}^{2}\right] \\
&=& \sum_p \omega_p \eta_p^{\dagger} \eta_p
\end{eqnarray}
where $\tilde{\phi} = \frac{\phi}{\sqrt{K}},\tilde{\theta} = \sqrt{K}\theta$ 
and $\omega_p = \sqrt{p^2 u^2 + m^2}$.

The calculations mirror that of the previous section. Writing $\phi = \phi^a + \delta \phi, \theta =
\theta^a + \delta \theta$, we find for a time $t$ after the quench
\begin{eqnarray}
&&\phi^a(x,t)= -\frac{i\pi}{L}\sum_{p\neq0}\left(\frac{L|p|}{2\pi}\right)^{1/2}\nonumber\\
&&\times \frac{1}{p} 
e^{-\alpha|p|/2-ipx}\left[a_{p}^{\dagger}
\left(\cos(\omega_p t) + i K \frac{u|p|}{\omega_p}\sin(\omega_pt)\right) 
\right. \nonumber \\
&&\left. + a_{-p}\left(\cos(\omega_p t) - i K \frac{u|p|}{\omega_p}
\sin(\omega_pt)\right)  \right], \\
&&\theta^a(x,t)= \frac{i\pi}{L}\sum_{p\neq0}\left(\frac{L|p|}{2\pi}\right)^{1/2} \nonumber \\
&&\times \frac{1}{|p|}e^{-\alpha|p|/2-ipx}\left[a_{p}^{\dagger}
\left(\cos(\omega_p t) + i\frac{\omega_p}{Ku|p|}
\sin(\omega_p t)\right) 
\right. \nonumber \\
&&\left. - a_{-p}
\left(\cos(\omega_p t) - i \frac{\omega_p}{K u |p|}
\sin(\omega_pt)\right)  
\right]
\end{eqnarray}
and
\begin{eqnarray}
\delta\phi(x,t) & = & \frac{i}{\pi v_{F}}\int_{0}^{\infty}\frac{dp}{p}
\cos[\omega_p t]\cos[px]h_{p}, \label{dphi1}\\
\delta\theta(x,t)  & = & \frac{-i}{\pi K v_{F}}\int_{0}^{\infty}\frac{dp}{p}\sin[\omega_p t]
\sin[px]\left(\frac{\omega_p}{u|p|}\right)h_{p} \nonumber \\\label{dtheta1}
\end{eqnarray}

\subsubsection{Magnetization after the quench}

The magnetization at a time $t$ after the quench is given by $\langle S^z(x,t)\rangle 
= -\frac{1}{\pi}\partial_x 
\phi(x,t)+ \frac{(-1)^x}{\pi\alpha}\langle \cos(2\phi(x,t))\rangle$. Due to the
presence of infra-red divergences, we find that for all times after the quench, 
\begin{eqnarray}
\left\langle e^{-2i\phi(x,t)}\right\rangle \sim 0
\end{eqnarray}
so that Ising order never develops.  
This is consistent with the results of~\cite{cazalilla} and 
may be an artifact of the semi-classical approximation.
In section~\ref{meanfield} we will compare this result with a mean-field treatment for the
order-parameter dynamics.

\begin{figure}
\begin{center}
\includegraphics[totalheight=22cm,width=9cm]{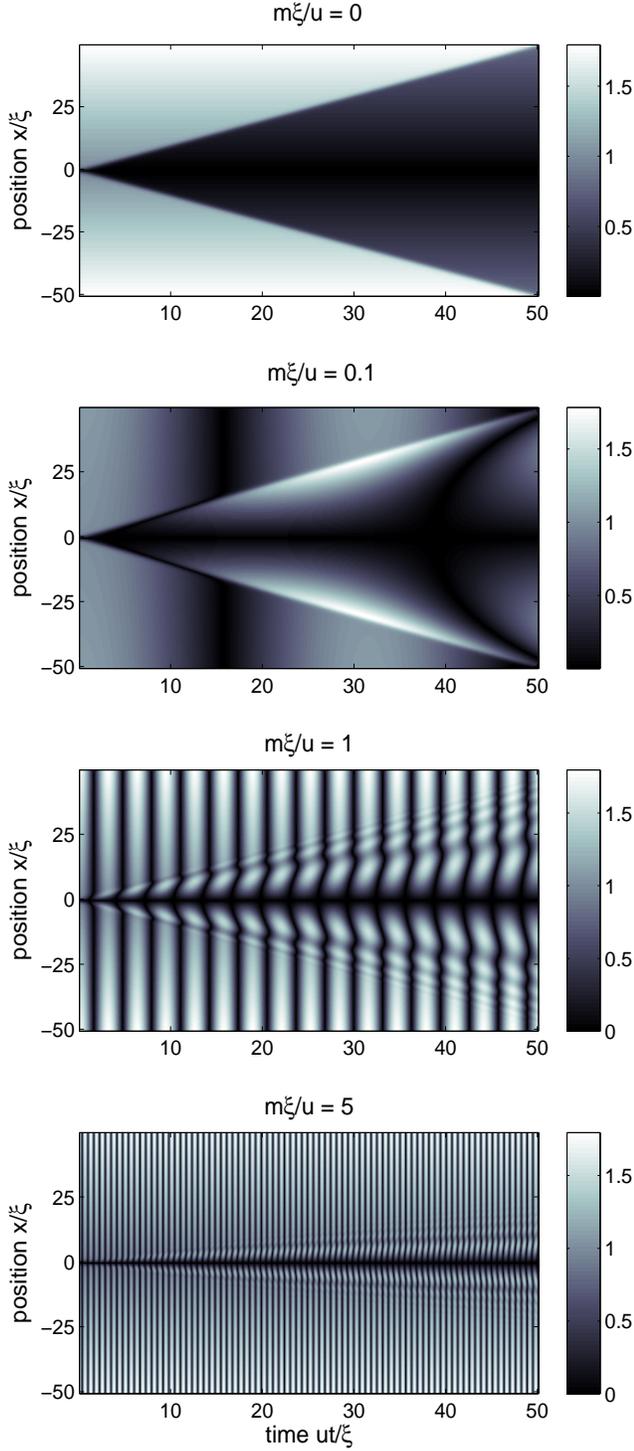}
\caption{\label{fig:scquench} Contour plot for $|S_z|$ after a 
``semi-classical'' quench for several different masses.}\label{Sztgap}
\end{center}
\end{figure}

Thus, within the validity of the semiclassical approximation we find the following time 
evolution for the initial domain wall profile,
\begin{eqnarray}
&&\left\langle S^{z}(x,t) \right\rangle = -\frac{1}{\pi}\partial_x 
\left(\delta \phi(x,t)\right) \nonumber \\
&&=\frac{h_{0}}{\pi v_F}\int_{0}^{\infty} dz
\cos\left(\frac{ut}{\xi}\sqrt{z^{2}+\left(\frac{m\xi}{u}\right)^{2}}\right)\nonumber \\
&&\times \frac{\sin\left(\frac{x}{\xi}z\right)}{\sinh\left(\frac{\pi z}{2}\right)}\label{Szgap}
\end{eqnarray}
The above is plotted in Fig.~\ref{Sztgap} for several
different masses. The domain wall profile shows a more complex time
evolution than for the quench in the gapless phase where a simple ballistic broadening
of the domain wall occurs. 
In the Ising phase instead we find oscillations of the local magnetization at 
the frequency of the mass. In addition
the domain wall spreads out less and less as the mass increases, a result which is in agreement
with a t-DMRG study of the XXZ chain~\cite{tdmrg05,tdmrg09}. Note however that we find oscillations
of the magnetization outside the light-cone at the time-scale of $1/m$. This is due to the nature
of the Klein-Gordon equation $\partial_t^2 -\partial_x^2 + m^2$ which has solutions of the form 
$e^{i k x-i \sqrt{k^2 + m^2}t}\xrightarrow{mt\gg 1} e^{-i m t}e^{i k x - i\frac{k^2}{2m}t}$.
The t-DMRG results on the other hand found oscillations only within the light cone.  

\begin{figure}
\begin{center}
\includegraphics[totalheight=7cm,width=7cm]{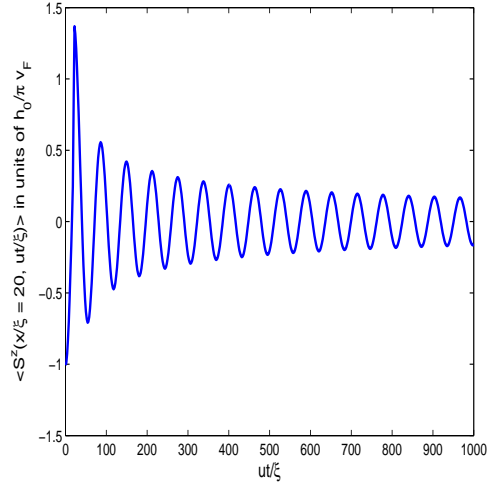}
\caption{\label{fig:scquench2}(Color online) The magnetization decays as $t^{-1/2}\,$ with oscillations of 
period $\sim m^{-1}$. Here, $m\xi/u = 0.1$.}\label{Sztgap1}
\end{center}
\end{figure}

We also find that the magnetization eventually
equilibrates at long times. Fig.~\ref{Sztgap1} shows the time evolution 
for the magnetization for a given point in position space. 
At long times Eq.~\ref{Szgap} may be evaluated in a stationary phase approximation 
which gives,
\begin{eqnarray}
\langle S^z(x,t) \rangle \simeq \frac{2h_0x}{\pi v_F}
\sqrt{\frac{2m}{\pi u^2 t}}\cos\left(mt + \frac{\pi}{4}\right)
\end{eqnarray}
showing that the magnetization decays locally as a power-law.

\subsubsection{Correlation functions after the quench}

Here we give results for some typical spin correlation functions. Again, due to the 
presence of infra-red divergences, the 
transverse-spin correlation function is found to vanish at all times after the
quench, 
\begin{eqnarray}
\langle e^{i\theta(x,t)} e^{-i\theta(0,t)}\rangle \sim 0 
\end{eqnarray}
On the other hand the $z-z$ correlation function is found to be
\begin{eqnarray}
\left\langle e^{2i\phi(x,t)}e^{-2i\phi(0,t)}\right\rangle = e^{2i(\delta\phi(x,t) -\delta\phi(x,0))}
A'(\alpha)\left(\frac{1}{x}\right)
\end{eqnarray}
For a purely interaction quench ($\delta\phi=0$) we find a power law decay in 
the Ising phase with an exponent which is half of that in the initial gapless phase, 
a result which was also found by~\cite{cazalilla} for homogeneous quenches.
The initial domain profile superimposes the additional structure
\begin{eqnarray}
e^{2i(\delta\phi(x,t) -\delta\phi(y,0))} \xrightarrow{mt \gg 1} e^{2i 
\left[\frac{h_{0}(x^2 - y^2)}{v_{F}}\sqrt{\frac{2m}{\pi u^{2} t}}
\cos\left(mt + \frac{\pi}{4}\right)\right]} 
\end{eqnarray}  
Thus at long times, the spatial oscillations due to the initial domain-wall eventually decay to zero.

\section{Effect of quenching a magnetic field and interactions: Mean-field treatment} \label{meanfield}

In this section we will use a mean-field argument to explore whether after quenching both
a magnetic field as well as interactions, local anti-ferromagnetic (AF) Ising order can develop. 
Thus our initial wave-function corresponds to the ground state of the following Hamiltonian
\begin{eqnarray}
H_i = J\sum_{j}\left[S_{j}^{x}S_{j+1}^{x} +  
S_{j}^{y}S_{j+1}^{y} +  \frac{h_j}{J}S_{j}^{z} \right]
\end{eqnarray}
with $J >0$, and we will assume $h_j = {\cal F} j a$.
While, as a result of the quench the wave-function evolves in time due to
\begin{eqnarray}
H_f = J\sum_{j}\left(S_{j}^{x}S_{j+1}^{x} +  
S_{j}^{y}S_{j+1}^{y}\right) + J^z \sum_{j}S_{j}^{z}S_{j+1}^{z} \label{Hfaf}
\end{eqnarray}
Performing a Jordan-Wigner transformation followed by the change of variables $c_i \rightarrow (-1)^i c_i$,
$H_i$ becomes identical to the Hamiltonian (Eq.~\ref{Hxxjw}) already studied in 
section~\ref{DWstate}, whereas the
Hamiltonian after the quench is, 
\begin{eqnarray}
&&H_f = -\frac{J}{2}\sum_i \left(c_i^{\dagger} c_{i+1} + h.c.\right) \nonumber \\
&&+ J^z\sum_i
\left(c_i^{\dagger}c_{i}
-\frac{1}{2}\right)\left(c_{i+1}^{\dagger}c_{i+1}-\frac{1}{2}\right) \label{HfAF}
\end{eqnarray}
Thus the initial state which is the ground state of $H_i$ (given by Eq.~\ref{eq:initstate}) 
evolves in time according to Eq.~\ref{HfAF}. In what follows we will study this time evolution using a 
mean-field approximation.

Defining the AF order-parameter at a time $t$, $\Delta_0(t)$ as $\langle c_i^{\dagger}(t) 
c_i(t)\rangle = \frac{1}{2}
+ (-1)^i \Delta_0(t)$, Eq.~\ref{HfAF} within a mean-field approximation becomes
\begin{eqnarray}
&&H_{f}^{mf} = \sum_{|k|<\pi/2} \epsilon_k \left(c_k^{\dagger} c_k 
- c^{\dagger}_{k+\pi} c_{k+\pi}\right) \nonumber \\
&&- 2J^z \Delta_0(t)
\sum_{|k|<\pi/2}\left(c_k^{\dagger} c_{k+\pi} + c_{k+\pi}^{\dagger} c_{k} \right) \label{HfAFmf}
\end{eqnarray}
where $\epsilon_k = -J \cos{ka}$, and $\Delta_0(t)$ is determined from the self-consistency
condition
\begin{eqnarray}
\Delta_0(t) = \frac{1}{N}\langle \Psi|e^{i H^{mf}_f t} \sum_i (-1)^i c_{i}^{\dagger} c_i e^{-i H^{mf}_ft}
| \Psi \rangle \label{mfsc}
\end{eqnarray}
where $N$ is the total number of sites. 

In terms of the amplitudes $\tilde{c}_{k\pm}$ of being in the adiabatic eigenstates of
$H^{mf}_f$, the self-consistency condition becomes
\begin{eqnarray}
&&\Delta_0(t) = \frac{1}{2N}\sum_{|k|<\pi/2} \left[
\sin2\theta_k(t)\left(|\tilde{c}_{k+}(t)|^2 - |\tilde{c}_{k-}(t)|^2\right)\right.
\nonumber \\
&&\left. -\cos2\theta_k(t)\left(\tilde{c}_{k+}^*(t)\tilde{c}_{k-}(t) 
+ \tilde{c}_{k-}^*(t)\tilde{c}_{k+}(t) \right)
\times \right. \nonumber \\
&&\left. \cos(\int_0^t dx (E_{k+}(x) - E_{k-}(x))) \right. \nonumber \\ 
&&\left. -i \cos2\theta_k(t)\left(\tilde{c}_{k+}^*(t)\tilde{c}_{k-}(t) 
- \tilde{c}_{k-}^*(t)\tilde{c}_{k+}(t) \right)
\times \right. \nonumber \\
&&\left. \sin(\int_0^t dx (E_{k+}(x) - E_{k-}(x))) \right] \label{Deltamf}
\end{eqnarray}
where  
\begin{eqnarray}
E_{k\pm}(t) = \pm\sqrt{\epsilon_k^2 + 4 J_z^2 \Delta_0^2(t)}\\
\tan{2\theta_k}(t) = -\frac{2J_z \Delta_0(t)}{\epsilon_k}
\end{eqnarray}
and the coefficients $\tilde{c}_{k\pm}$ satisfy the equations of motion,
\begin{eqnarray}
\frac{d\tilde{c}_{k\pm}(t)}{d t} = \tilde{c}_{k\mp}(t)
\left[\frac{\pm\epsilon_k J_z \frac{d\Delta_0}{dt}}{\epsilon_k^2 + 4 J_z^2\Delta_0^2}\right] 
e^{\pm i\int_0^t dx
(E_{k+}(x) - E_{k-}(x))}\nonumber \\
\end{eqnarray}

In an adiabatic approximation $\frac{\dot{\Delta}_0}{\Delta^2_0} \ll 1$, we can approximate 
$\tilde{c}_{k\pm}$
by its value at $t=0$. In this case Eq.~\ref{Deltamf} becomes
\begin{eqnarray}
&&\Delta_0 \simeq \frac{1}{2N}\sum_{|k|<\pi/2} \left[
\sin2\theta_k(t)\left(\sin{2\theta_k}(t) I_{k}^{x} + \cos{2\theta_k}(t) I_{k}^{z}\right)\right.
\nonumber \\
&&\left. -\cos2\theta_k(t)\left(\sin{2\theta_k}(t)I_{k}^{z} -\cos{2\theta_k}(t) I_{k}^{x}\right)
\times \right. \nonumber \\
&&\left. \cos(\int_0^t dx (E_{k+}(x) - E_{k-}(x))) \right. \nonumber \\ 
&&\left. -i \cos2\theta_k(t)I_{k}^{y}\sin(\int_0^t dx (E_{k+}(x) - E_{k-}(x))) \right]
\label{Deltamfad}
\end{eqnarray}
where $I_{k}^{x,y,z}$ denote the following expectation values in the initial state before the quench,
\begin{eqnarray}
I_{k}^{x} = \langle\Psi| c_k^{\dagger}c_{k+\pi} + c_{k+\pi}^{\dagger} c_{k} | \Psi \rangle \\
I_{k}^{z} = \langle\Psi| c_k^{\dagger}c_{k} - c_{k+\pi}^{\dagger} c_{k+\pi} | \Psi \rangle \\
I_{k}^{y} = \langle\Psi|c_{k+\pi}^{\dagger} c_{k} - c_k^{\dagger}c_{k+\pi} | \Psi \rangle
\end{eqnarray}
Since in the initial domain wall state, $I_{k}^{x,y,z}\sim 0$, this implies that
$\Delta_0(t)=0$, and hence anti-ferromagnetic order does not develop within a
mean-field and adiabatic approximation. This result can also be understood from 
the fact that the initial state is a highly excited state of the 
anti-ferromagnetic model as it corresponds to 
introducing a domain wall in the staggered magnetization at almost every site~\cite{tdmrg05}. 

The above conclusion also holds if we relax the adiabatic approximation. To see this note that 
at every $k$, the mean-field Hamiltonian Eq.~\ref{HfAFmf} corresponds to a pseudo-spin $1/2$
$\vec{\tau}_k$ where~\cite{Demler09b} $\tau_{k}^z = c_k^{\dagger}c_{k} - c_{k+\pi}^{\dagger} c_{k+\pi}$,
$\tau_{k}^x = c_k^{\dagger}c_{k+\pi} + c_{k+\pi}^{\dagger} c_{k}$ and 
$\tau_k^y = i\left(c_{k+\pi}^{\dagger} c_{k} - c_k^{\dagger}c_{k+\pi}\right)$.
This pseudo-spin is subjected to a magnetic field that has to be determined 
self-consistently at each time $t$ where for $t=0$ 
$\langle \tau_{k}^{x,z}(0) \rangle = I_{k}^{x,z}, \langle \tau_{k}^{y}(0) \rangle = i I_{k}^{y}$.
Since for the initial domain wall state $\langle \vec{\tau}(0)\rangle= 0$,
it implies $\frac{d^n\langle \tau_{k}^x\rangle}{dt^n}(t=0) = 0$ to all orders in $n$, so that 
anti-ferromagnetic order ($\propto \sum_k \langle \tau_k^x\rangle$) will
not develop.

\section{Conclusions} \label{conclusions}

In summary we have studied quenched dynamics in a XXZ chain starting from an initial inhomogeneous 
state where the spins are arranged in a domain wall profile. The dynamics of the domain wall is
found to be qualitatively different within the gapless XX phase and the gapped Ising phase. In the
former the domain wall broadens ballistically, while in the latter the domain wall spreads out
less and less the deeper one is in the Ising phase. The magnetization is locally found to oscillate,
and eventually decays as a power-law. Although the results in the Ising phase were obtained within
a semiclassical treatment, they are in qualitative agreement with t-DMRG simulations of the
XXZ chain~\cite{tdmrg05,tdmrg09}. 

We have also presented results for the time-evolution of two-point correlation functions. 
For quenches within the gapless XX phase we find that
the longitudinal spin correlation function ($C^{zz}$) reaches a steady state which is
independent of the initial domain wall. On the other hand, 
the transverse spin correlation function ($C^{xx}$) reaches a nonequilibrium steady state which
retains memory of the initial domain wall. This memory appears as 
an inhomogeneous spin wave pattern with a wavelength which is inversely related to the 
height of the initial domain wall. This result was obtained by using three
different methods: exact solution of the fermionic problem (Eqns.~\ref{Cxxansgen},~\ref{lamgen}), 
bosonization (Eqns~\ref{S+-int},~\ref{lamint}) and 
the solution of the classical equations of motions for spins (Eq.~\ref{S+class}).  
The spatial oscillations are found to arise due
to the dephasing of transverse spin components as the domain wall broadens. 
We also find that a Gibbs
ensemble argument is not adequate in capturing the spin wave pattern.

For quenches into the gapped Ising phase, all inhomogeneities both in the
local magnetization and in the two-point correlation functions are found to eventually decay away. This
may be an artifact of the semiclassical approximation which neglects soliton creation.~\cite{cazalilla}
We would also like to point out a recent preprint~\cite{Caux10}
that studies time-evolution of an initial domain wall in the gapped Ising phase using 
Algebraic Bethe Ansatz. The authors study the Loschmidt echo and find a lack of thermalization.

Finally we perform a mean-field treatment for the time-evolution of the 
anti-ferromagnetic Ising gap and find that  
for an initial state corresponding to a domain wall profile, anti-ferromagnetic 
order never develops. This should be expected on the grounds 
that the initial domain wall profile corresponds to a highly excited state of the 
anti-ferromagnetic Ising model.

\begin{acknowledgments}

The authors are particularly indebted to Eugene Demler and Ulrich Schollw\"{o}ck for helpful discussions.
This work was supported by NSF-DMR (Award No. 0705584).

\end{acknowledgments}

\appendix

\section{Diagonalization of the $XX$-model} \label{diagonol}

The Hamiltonian for the $XX$ model is
\begin{equation}
H_{xx} = -J\sum_{j}\left[ S_{j}^{x}S_{j+1}^{x} +  
S_{j}^{y}S_{j+1}^{y} \right].
\end{equation}
The Jordan-Wigner transformation maps
the spin Hamiltonian to that for spin-less fermions $c_{j}\,$ and $c_{j}^{\dagger}$ 
which are defined 
in terms of the spin raising and lowering operators $S_{j}^{\pm} = S_{j}^{x} \pm i S_{j}^{y}$ as follows,
\begin{eqnarray}
S_{j}^{-} & = & \exp\left[-i\pi\sum_{n=-N/2 + 1}^{j-1}c^{\dagger}_{n}c_{n}\right]c_{j},\\
S_{j}^{+} & = & \exp\left[i\pi\sum_{n=-N/2+1}^{j-1}c^{\dagger}_{n}c_{n}\right]c^{\dagger}_{j},
\end{eqnarray}
The Hamiltonian is diagonal in momentum space $c_j =\frac{1}{\sqrt{N}}\sum_k c_k e^{ikj}$ where,
\begin{eqnarray}
&&H_{xx} = \sum_{k}\epsilon_{k}c_{k}^{\dagger}c_{k}  \\
&&\epsilon_{k} = -J\cos k.
\end{eqnarray}
Thus the operators $c_{k}$, have the trivial time-evolution $c_k(t) = c_k(0)e^{-i\epsilon_k t}$.
In terms of the $\eta_{m}\,$ quasi-particles that
diagonalize the Wanner-Stark problem (Eq.~\ref{Hxxjw}), 
the transformation given by Eqns.~(\ref{eq:etaj},~\ref{eq:etaj1})
yields
\begin{eqnarray}
c_{j}(t) & = & \frac{1}{N}\sum_{k,m}e^{ik(j-m)-i\alpha\sin k}e^{-i\epsilon_{k}t}\eta_{m}\label{cjtim1}\\
c^{\dagger}_{j}(t) & = & \frac{1}{N}
\sum_{k,m}e^{-ik(j-m)+i\alpha\sin k}e^{i\epsilon_{k}t}\eta^{\dagger}_{m}\label{cjtim2}
\end{eqnarray}

\section{Time evolution in the $XX$ model starting from initial domain walls of different heights} 
\label{m0vary}

Antal {\it et al.} have considered more general domain walls~\cite{antal} 
in which the homogeneous magnetization on either side of the wall is taken to be $\pm m_{0}$, 
where $0<m_{0} < 1/2$. They have studied the time evolution of the domain wall 
under the influence of the $XX$ hamiltonian.
In this section we will extend their results by studying how the transverse spin
correlation function evolves in time. 

A domain wall of height $m_0$ may be constructed as follows~\cite{antal},
\begin{equation}
\left|\Psi_{m_{0}}\right\rangle  = \prod_{k=-k_{-}}^{k_{-}}
R_{k}^{\dagger}\prod_{k=-k_{+}}^{k_{+}}L_{k}^{\dagger}\left|0\right\rangle,
\end{equation}
where 
\begin{eqnarray}
R_{k} & = & \frac{1}{\sqrt{N}}\sum_{j>0}e^{-ikj}c_{j},\\
L_{k} & = & \frac{1}{\sqrt{N}}\sum_{j\leq0}e^{-ikj}c_{j},
\end{eqnarray}
and $k_{\pm} = \pi\left(\frac{1}{2}\pm m_{0}\right)$. 
Due to the XX Hamiltonian $H= -J\sum_k\cos(k)c_k^{\dagger}c_k$, the operators $c_{j}\,$ evolve as 
\begin{equation}
c_{j}(t) = \sum_{m}i^{m-j}J_{m-j}(Jt)c_{m}(0),
\end{equation}
In order to compute any correlation function, the contraction we need is
\begin{eqnarray}
&&\left\langle c_{j}^{\dagger}(t)c_{j+n}(t)\right\rangle = 
\sum_{s,m}i^{s-m-n}J_{m-j}(Jt)J_{s-j-n}(Jt)\nonumber \\
&&\times \left\langle c_{m}^{\dagger}(0)c_{s}(0)\right\rangle.
\end{eqnarray}
where, 
\begin{equation}
\left\langle c_{s}^{\dagger}(0)c_{m}(0)\right\rangle = 
\left\{\begin{array}{cc} 0 & \mbox{if }s\cdot m < 0\\
\frac{\sin\left[\left(\frac{\pi}{2}+\pi m_{0}\right)\left(s-m\right)\right]}
{\pi(s-m)} & \mbox{if } s,m <0\\
\frac{\sin\left[\left(\frac{\pi}{2}-\pi m_{0}\right)\left(s-m\right)\right]}{\pi(s-m)} & \mbox{if } s,m >0
\end{array}\right.
\end{equation}
Thus,
\begin{eqnarray}
&&\left\langle c_{j}^{\dagger}(t)c_{j+n}(t)\right\rangle =\nonumber \\ 
&&\sum_{m,s\leq0}
i^{s-m-n}J_{m-j}(Jt)J_{s-j-n}(Jt)\nonumber\\
&&\times \frac{\sin\left[\left(\frac{\pi}{2}+\pi m_{0}
\right)(m-s)\right]}{\pi(m-s)} \nonumber\\
&& + \sum_{m,s>0}i^{s-m-n}J_{m-j}(Jt)J_{s-j-n}(Jt)\nonumber \\
&&\times \frac{\sin\left[\left(\frac{\pi}{2}-\pi m_{0}\right)(m-s)\right]}{\pi(m-s)}.
\end{eqnarray}
We write both terms as sums over positive $m$, $s$, 
(i.e., letting $m,s\rightarrow-m,-s\,$ in the second term), 
and rewrite $\sum_{m,s>0} \rightarrow \sum_{m>0}\sum_{l=-\infty}^{\infty}$
with $l = s-m$, and use the identity
\begin{eqnarray}
&&\sum_{m>0}J_{j+m}(x)J_{j+n+m}(x) = \frac{x}{2n}\left(J_{j+1}(x)J_{j+n}(x) \right. \nonumber\\
&&\left. - J_{j}(x)J_{j+n+1}(x)\right).
\end{eqnarray}
Next we employ the asymptotic expansion for large arguments, 
$J_{n}(x) \sim \sqrt{\frac{2}{\pi x}}\cos\left[x-\frac{\pi}{2}n-\frac{\pi}{4}\right]$ to find
the following nonequilibrium steady state expressions,
\begin{eqnarray}
&&\lim_{Jt\rightarrow\infty}\left\langle c_{j}^{\dagger}(t)c_{j+n}(t)\right\rangle \simeq\nonumber\\
&& \sum_{l=-\infty}^{\infty}\left[i^{l-n}\frac{\sin\left[\left(\frac{\pi}{2}
-\pi m_{0}\right)l\right]\sin\left[\frac{\pi}{2}\left(l-n\right)\right]}{\pi^{2}l(l-n)} 
\right. \nonumber\\
&&\left. + i^{l+n}\frac{\sin\left[\left(\frac{\pi}{2}+\pi m_{0}\right)l\right]\sin\left[\frac{\pi}{2}
\left(l+n\right)\right]}{\pi^{2}l(l+n)}\right]\label{asmgen}.
\end{eqnarray}
Examining the limit of $m_{0} = \frac{1}{2}$, we recover  
exactly the values in Eq.~\ref{asm1},~\ref{asm2}. We use Eq.~\ref{asmgen} in the evaluation of the
determinant in Eq.~\ref{detC} required for computing the transverse spin correlation function. 
We recover a spatial period of oscillation in $C_{xx}(j,j+n, t\rightarrow \infty)$ 
that increases with smaller $m_{0}$ (see Fig.~\ref{fig:cxxperiod}).


\begin{thebibliography}{5}
\bibitem{zeromode} Note that in the limit of a large number of sites, the result is insensitive
to whether the $m=0$ mode is included in the sum or not.
\end{thebibliography}
\end{document}